\pdfoutput=1

\documentclass[12pt,a4paper]{article}

\usepackage{ifthen} 
\newboolean{pdflatex}
\setboolean{pdflatex}{true} 

\newboolean{articletitles}
\setboolean{articletitles}{true} 

\newboolean{uprightparticles}
\setboolean{uprightparticles}{false} 

\newboolean{inbibliography}
\setboolean{inbibliography}{false} 

\def\paperauthors{LHCb collaboration} 
\def\paperasciititle{Measurement of the CKM angle gamma using B->DK with D->K0SPiPi and D->K0SKK decays} 
\def\papertitle{Measurement of the CKM angle $\gamma$ \\ 
using \BtoDK with \\ 
$D \to \KsPiPi, \KsKK$ decays} 
\def\paperkeywords{{High Energy Physics}, {LHCb}} 
\def\papercopyright{\the\year\ CERN for the benefit of the LHCb collaboration}
\def\paperlicence{CC-BY-4.0 licence}
\def\paperlicenceurl{https://creativecommons.org/licenses/by/4.0/}

\usepackage[top=1in, bottom=1.25in, left=1in, right=1in]{geometry}

\usepackage{microtype}
\usepackage{lineno}  
\usepackage{xspace} 
\usepackage{caption} 

\usepackage{graphicx}  
\usepackage{color}
\usepackage{colortbl}
\graphicspath{{./figs/}} 

\usepackage{amsmath} 
\usepackage{amssymb}
\usepackage{amsfonts}
\usepackage{upgreek} 

\usepackage{tabularx}
\newcolumntype{Y}{>{\centering\arraybackslash}X}
\newcommand*\patchAmsMathEnvironmentForLineno[1]{%
\expandafter\let\csname old#1\expandafter\endcsname\csname #1\endcsname
\expandafter\let\csname oldend#1\expandafter\endcsname\csname
end#1\endcsname
 \renewenvironment{#1}%
   {\linenomath\csname old#1\endcsname}%
   {\csname oldend#1\endcsname\endlinenomath}%
}
\newcommand*\patchBothAmsMathEnvironmentsForLineno[1]{%
  \patchAmsMathEnvironmentForLineno{#1}%
  \patchAmsMathEnvironmentForLineno{#1*}%
}
\AtBeginDocument{%
\patchBothAmsMathEnvironmentsForLineno{equation}%
\patchBothAmsMathEnvironmentsForLineno{align}%
\patchBothAmsMathEnvironmentsForLineno{flalign}%
\patchBothAmsMathEnvironmentsForLineno{alignat}%
\patchBothAmsMathEnvironmentsForLineno{gather}%
\patchBothAmsMathEnvironmentsForLineno{multline}%
\patchBothAmsMathEnvironmentsForLineno{eqnarray}%
}

\usepackage{hyperxmp}

\usepackage[pdftex,
            pdfauthor={\paperauthors},
            pdftitle={\paperasciititle},
            pdfkeywords={\paperkeywords},
            pdfcopyright={Copyright (C) \papercopyright},
            pdflicenseurl={\paperlicenceurl}]{hyperref}

\usepackage[all]{hypcap} 

\usepackage{xspace} 
\usepackage{upgreek}

\def\lhcb {\mbox{LHCb}\xspace}

\def\babar  {\mbox{BaBar}\xspace}
\def\belle  {\mbox{Belle}\xspace}

\def\MagUp {\mbox{\em Mag\kern -0.05em Up}\xspace}

\ifthenelse{\boolean{uprightparticles}}%
{

 \def\Ppi         {\ensuremath{\uppi}\xspace}

 \def\PDelta      {\ensuremath{\Delta}\xspace}                 
 \def\PXi      {\ensuremath{\Xi}\xspace}                 
 \def\PLambda      {\ensuremath{\Lambda}\xspace}                 
 \def\PSigma      {\ensuremath{\Sigma}\xspace}                 
 \def\POmega      {\ensuremath{\Omega}\xspace}                 
 \def\PUpsilon      {\ensuremath{\Upsilon}\xspace}                 
 
 \def\PB      {\ensuremath{\mathrm{B}}\xspace}                 
                  
 \def\PD      {\ensuremath{\mathrm{D}}\xspace}

 \def\PK      {\ensuremath{\mathrm{K}}\xspace}

 \def\Pb      {\ensuremath{\mathrm{b}}\xspace}                 
 \def\Pc      {\ensuremath{\mathrm{c}}\xspace}

 \def\Ph      {\ensuremath{\mathrm{h}}\xspace}                 
 \def\Pi      {\ensuremath{\mathrm{i}}\xspace}

 \def\Ps      {\ensuremath{\mathrm{s}}\xspace}

}
{

 \def\Ppi         {\ensuremath{\pi}\xspace}

 \mathchardef\PDelta="7101
 \mathchardef\PXi="7104
 \mathchardef\PLambda="7103
 \mathchardef\PSigma="7106
 \mathchardef\POmega="710A
 \mathchardef\PUpsilon="7107
                  
 \def\PB      {\ensuremath{B}\xspace}                 
                  
 \def\PD      {\ensuremath{D}\xspace}

 \def\PK      {\ensuremath{K}\xspace}

 \def\Pb      {\ensuremath{b}\xspace}                 
 \def\Pc      {\ensuremath{c}\xspace}

 \def\Ph      {\ensuremath{h}\xspace}                 
 \def\Pi      {\ensuremath{i}\xspace}

 \def\Ps      {\ensuremath{s}\xspace}

}

\makeatletter
\ifcase \@ptsize \relax
  \newcommand{\miniscule}{\@setfontsize\miniscule{4}{5}}
\or
  \newcommand{\miniscule}{\@setfontsize\miniscule{5}{6}}
\or
  \newcommand{\miniscule}{\@setfontsize\miniscule{5}{6}}
\fi
\makeatother

\DeclareRobustCommand{\optbar}[1]{\shortstack{{\miniscule (\rule[.5ex]{1.25em}{.18mm})}
  \\ [-.7ex] $#1$}}

\def\squark    {{\ensuremath{\Ps}}\xspace}

\def\cquark    {{\ensuremath{\Pc}}\xspace}

\def\bquark    {{\ensuremath{\Pb}}\xspace}

\def\hadron {{\ensuremath{\Ph}}\xspace}
\def\pion   {{\ensuremath{\Ppi}}\xspace}

\def\pip    {{\ensuremath{\pion^+}}\xspace}
\def\pim    {{\ensuremath{\pion^-}}\xspace}
\def\pipm   {{\ensuremath{\pion^\pm}}\xspace}

\def\kaon    {{\ensuremath{\PK}}\xspace}
  \def\Kbar    {{\kern 0.2em\overline{\kern -0.2em \PK}{}}\xspace}

\def\KorKbar    {\kern 0.18em\optbar{\kern -0.18em K}{}\xspace}
\def\Kz      {{\ensuremath{\kaon^0}}\xspace}
\def\Kzb     {{\ensuremath{\Kbar{}^0}}\xspace}
\def\Kp      {{\ensuremath{\kaon^+}}\xspace}
\def\Km      {{\ensuremath{\kaon^-}}\xspace}
\def\Kpm     {{\ensuremath{\kaon^\pm}}\xspace}
\def\Kmp     {{\ensuremath{\kaon^\mp}}\xspace}
\def\KS      {{\ensuremath{\kaon^0_{\mathrm{ \scriptscriptstyle S}}}}\xspace}
\def\KL      {{\ensuremath{\kaon^0_{\mathrm{ \scriptscriptstyle L}}}}\xspace}
\def\Kstarz  {{\ensuremath{\kaon^{*0}}}\xspace}

\def\Kstar   {{\ensuremath{\kaon^*}}\xspace}

\def\Kstarpm {{\ensuremath{\kaon^{*\pm}}}\xspace}

  \def\Dbar    {{\kern 0.2em\overline{\kern -0.2em \PD}{}}\xspace}
\def\D       {{\ensuremath{\PD}}\xspace}

\def\DorDbar    {\kern 0.18em\optbar{\kern -0.18em D}{}\xspace}
\def\Dz      {{\ensuremath{\D^0}}\xspace}
\def\Dzb     {{\ensuremath{\Dbar{}^0}}\xspace}

\def\Dstar   {{\ensuremath{\D^*}}\xspace}

\def\Dstarm  {{\ensuremath{\D^{*-}}}\xspace}
\def\Dstarpm {{\ensuremath{\D^{*\pm}}}\xspace}

\def\B       {{\ensuremath{\PB}}\xspace}
\def\Bbar    {{\ensuremath{\kern 0.18em\overline{\kern -0.18em \PB}{}}}\xspace}

\def\BorBbar    {\kern 0.18em\optbar{\kern -0.18em B}{}\xspace}
\def\Bz      {{\ensuremath{\B^0}}\xspace}

\def\Bu      {{\ensuremath{\B^+}}\xspace}
\def\Bub     {{\ensuremath{\B^-}}\xspace}
\def\Bp      {{\ensuremath{\Bu}}\xspace}
\def\Bm      {{\ensuremath{\Bub}}\xspace}
\def\Bpm     {{\ensuremath{\B^\pm}}\xspace}

\def\Bs      {{\ensuremath{\B^0_\squark}}\xspace}
\def\Bsb     {{\ensuremath{\Bbar{}^0_\squark}}\xspace}

  \def\Y#1S{\ensuremath{\PUpsilon{(#1S)}}\xspace}

\def\Lbar        {{\ensuremath{\kern 0.1em\overline{\kern -0.1em\PLambda}}}\xspace}
\def\LorLbar    {\kern 0.18em\optbar{\kern -0.18em \PLambda}{}\xspace}

\def\to                 {\ensuremath{\rightarrow}\xspace}

\def\eps   {{\ensuremath{\varepsilon}}\xspace}

\def\CP                {{\ensuremath{C\!P}}\xspace}

\def\AT#1     {\ensuremath{A_{\mathrm{T}}^{#1}}\xspace}           

\def\C#1      {\ensuremath{\mathcal{C}_{#1}}\xspace}                       
\def\Cp#1     {\ensuremath{\mathcal{C}_{#1}^{'}}\xspace}                    
\def\Ceff#1   {\ensuremath{\mathcal{C}_{#1}^{\mathrm{(eff)}}}\xspace}        
\def\Cpeff#1  {\ensuremath{\mathcal{C}_{#1}^{'\mathrm{(eff)}}}\xspace}       
\def\Ope#1    {\ensuremath{\mathcal{O}_{#1}}\xspace}                       
\def\Opep#1   {\ensuremath{\mathcal{O}_{#1}^{'}}\xspace}                    

\newcommand{\tev}{\ifthenelse{\boolean{inbibliography}}{\ensuremath{~T\kern -0.05em eV}}{\ensuremath{\mathrm{\,Te\kern -0.1em V}}}\xspace}
\newcommand{\gev}{\ensuremath{\mathrm{\,Ge\kern -0.1em V}}\xspace}
\newcommand{\mev}{\ensuremath{\mathrm{\,Me\kern -0.1em V}}\xspace}
\newcommand{\kev}{\ensuremath{\mathrm{\,ke\kern -0.1em V}}\xspace}
\newcommand{\ev}{\ensuremath{\mathrm{\,e\kern -0.1em V}}\xspace}
\newcommand{\gevc}{\ensuremath{{\mathrm{\,Ge\kern -0.1em V\!/}c}}\xspace}
\newcommand{\mevc}{\ensuremath{{\mathrm{\,Me\kern -0.1em V\!/}c}}\xspace}
\newcommand{\gevcc}{\ensuremath{{\mathrm{\,Ge\kern -0.1em V\!/}c^2}}\xspace}
\newcommand{\gevgevcccc}{\ensuremath{{\mathrm{\,Ge\kern -0.1em V^2\!/}c^4}}\xspace}
\newcommand{\mevcc}{\ensuremath{{\mathrm{\,Me\kern -0.1em V\!/}c^2}}\xspace}

\def\mum  {\ensuremath{{\,\upmu\mathrm{m}}}\xspace}

\def\invfb   {\ensuremath{\mbox{\,fb}^{-1}}\xspace}

\newcommand{\chisq}{\ensuremath{\chi^2}\xspace}

\newcommand{\chisqip}{\ensuremath{\chi^2_{\text{IP}}}\xspace}

\def\gsim{{~\raise.15em\hbox{$>$}\kern-.85em
          \lower.35em\hbox{$\sim$}~}\xspace}
\def\lsim{{~\raise.15em\hbox{$<$}\kern-.85em
          \lower.35em\hbox{$\sim$}~}\xspace}

\def\sPlot{\mbox{\em sPlot}\xspace}

\def\ptot       {\mbox{$p$}\xspace}
\def\pt         {\mbox{$p_{\mathrm{ T}}$}\xspace}

\def\degrees{\ensuremath{^{\circ}}\xspace}

\def\evtgen     {\mbox{\textsc{EvtGen}}\xspace}

\def\geant      {\mbox{\textsc{Geant4}}\xspace}

\def\photos     {\mbox{\textsc{Photos}}\xspace}

\def\pythia     {\mbox{\textsc{Pythia}}\xspace}

\def\tell1  {TELL1\xspace}
\def\ukl1   {UKL1\xspace}

\newcommand{\BtoDpi}{\ensuremath{\Bpm\to\PD\pipm}\xspace}
\newcommand{\BtoDK}{\ensuremath{\Bpm\to\PD\Kpm}\xspace}

\newcommand{\Ks}{\ensuremath{\KS}\xspace}
\DeclareRobustCommand{\titleoptbar}[1]{\shortstack{{\miniscule (\rule[.5ex]{1.75em}{.18mm})}   \\ [-.3ex] $#1$}}
\DeclareRobustCommand{\titleoptbarlowcase}[1]{\shortstack{{\miniscule (\rule[.5ex]{1.25em}{.18mm})}   \\ [-.3ex] $#1$}}
\DeclareRobustCommand{\optbarlowcase}[1]{\shortstack{{\miniscule (\rule[.5ex]{1.0em}{.18mm})}   \\ [-.7ex] $#1$}}
\def\nuornubar{\optbarlowcase{\nu}{} \xspace}

\newcommand{\mupm}{\ensuremath{\mu^\pm}\xspace}
\newcommand{\BpmtoDpipm}{\ensuremath{\Bpm\to\PD\pipm}\xspace}
\newcommand{\BpmtoDKpm}{\ensuremath{\Bpm\to\PD\Kpm}\xspace}
\newcommand{\BpmtoDhpm}{\ensuremath{\Bpm\to\PD\Ph^\pm}\xspace}

\newcommand{\TitleBztoDstmu}{\ensuremath{\titleoptbar{B}{} \to\Dstarpm\mu^\mp\titleoptbarlowcase{\nu}\!_\mu}\xspace}
\newcommand{\BztoDstmu}{\ensuremath{\BorBbar\!\to\Dstarpm\mu^\mp\nuornubar_{\!\!\mu}}\xspace}

\newcommand{\BztoDstmuX}{\ensuremath{\BorBbar\!\to\Dstarpm\mu^\mp\nuornubar_{\!\!\mu} X}\xspace}

\newcommand{\BtoDKst}{\ensuremath{\Bz\to\PD\Kstarz}\xspace}

\newcommand{\msqmin}{\ensuremath{m^2_-}\xspace}
\newcommand{\msqplus}{\ensuremath{m^2_+}\xspace}

\newcommand{\KsPiPi}{\ensuremath{\KS\pip\pim}\xspace}
\newcommand{\KsKK}{\ensuremath{\KS\Kp\Km}\xspace}
\newcommand{\Kshh}{\ensuremath{\KS\Ph^+\Ph^-}\xspace}

\newcommand{\DtoKsPiPi}{\ensuremath{\PD\to\KsPiPi}\xspace}
\newcommand{\DtoKsKK}{\ensuremath{\PD\to\KsKK}\xspace}
\newcommand{\DtoKshh}{\ensuremath{\PD\to\Kshh}\xspace}

\newcommand{\etaDPi}{\ensuremath{\eta(m^2_-,m^2_+)_{D\pi}}\xspace}
\newcommand{\etaDst}{\ensuremath{\eta(m^2_-,m^2_+)_{\Dstar\!\mu}}\xspace}

\newcommand{\xpm}{\ensuremath{x_{\pm}}\xspace}
\newcommand{\ypm}{\ensuremath{y_{\pm}}\xspace}
\newcommand{\xp}{\ensuremath{x_{+}}\xspace}
\newcommand{\yp}{\ensuremath{y_{+}}\xspace}
\newcommand{\xm}{\ensuremath{x_{-}}\xspace}
\newcommand{\ym}{\ensuremath{y_{-}}\xspace}
\def\xy {\ensuremath{x_{\pm}},\xspace \ensuremath{y_{\pm}}\xspace}

\usepackage{cite} 
\usepackage{mciteplus}

\usepackage{longtable} 

\begin{document}

\renewcommand{\thefootnote}{\fnsymbol{footnote}}
\setcounter{footnote}{1}


\begin{titlepage}
\pagenumbering{roman}

\vspace*{-1.5cm}
\centerline{\large EUROPEAN ORGANIZATION FOR NUCLEAR RESEARCH (CERN)}
\vspace*{1.5cm}
\noindent
\begin{tabular*}{\linewidth}{lc@{\extracolsep{\fill}}r@{\extracolsep{0pt}}}
\ifthenelse{\boolean{pdflatex}}
{\vspace*{-1.5cm}\mbox{\!\!\!\includegraphics[width=.14\textwidth]{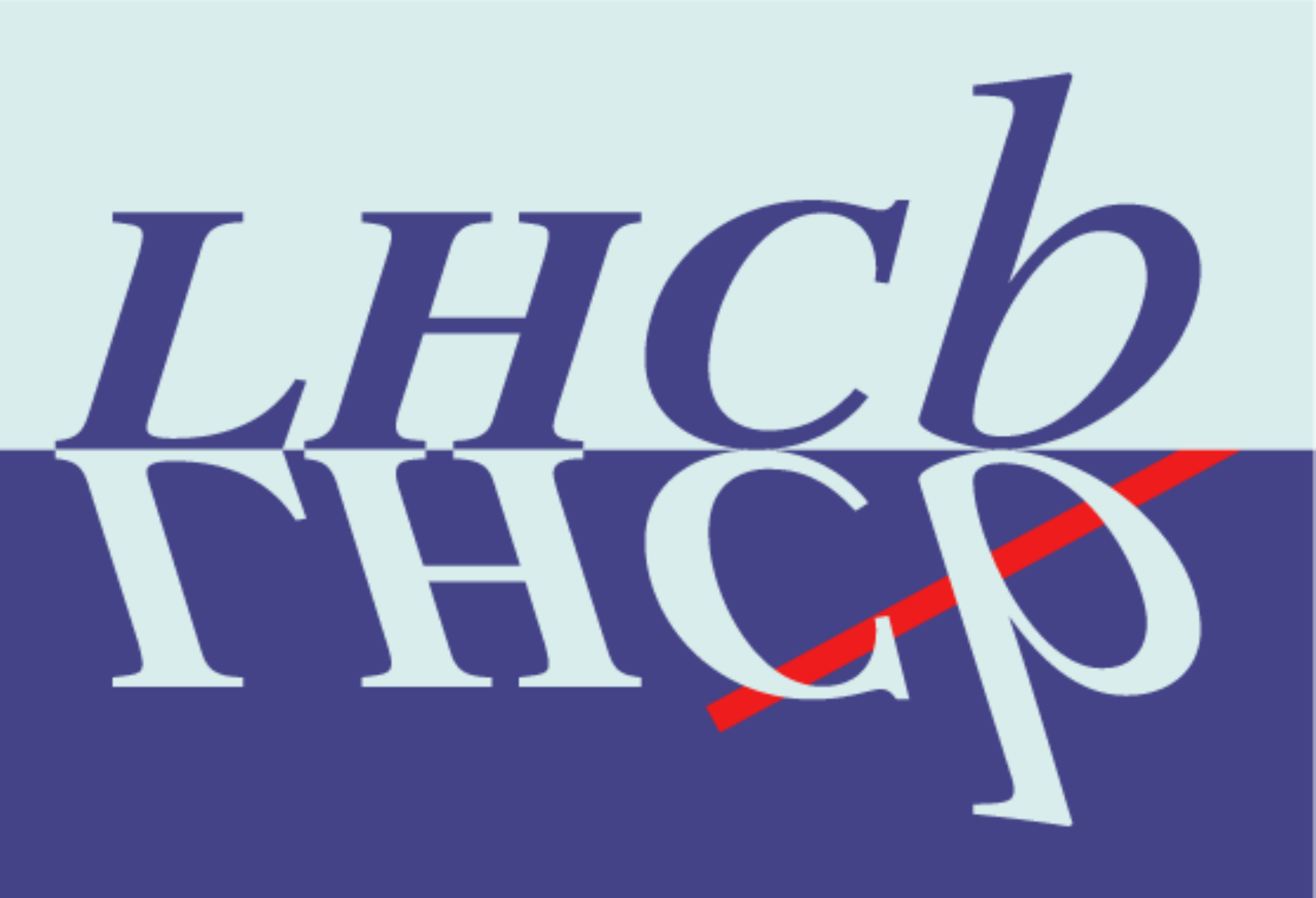}} & &}%
{\vspace*{-1.2cm}\mbox{\!\!\!\includegraphics[width=.12\textwidth]{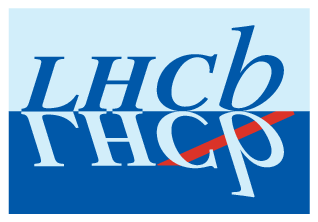}} & &}%
\\
 & & CERN-EP-2018-135 \\  
 & & LHCb-PAPER-2018-017 \\  
 & & 21 September 2018 \\ 
 & &  \\
\end{tabular*}

\vspace*{2.0cm}

{\normalfont\bfseries\boldmath\huge
\begin{center}
  \papertitle 
\end{center}
}

\vspace*{0.5cm}

\begin{center}
\paperauthors\footnote{Authors are listed at the end of this paper.}
\end{center}

\vspace{\fill}

\begin{abstract}
  \noindent
A binned Dalitz plot analysis of $B^\pm \to D K^\pm$ decays, with
\DtoKsPiPi and \DtoKsKK, is used to perform a measurement of the \CP-violating observables $x_{\pm}$ and $y_{\pm}$, which are sensitive to the Cabibbo-Kobayashi-Maskawa angle $\gamma$.
The analysis is performed without assuming any $D$ decay
model, through the use of information on the strong-phase variation over the Dalitz plot 
from the CLEO collaboration.
Using a sample of proton-proton collision data collected with the LHCb experiment in 2015 and 2016,
and corresponding to an integrated luminosity of 2.0\invfb,
the values of the \CP violation parameters are found to be
$x_-  = (  9.0 \pm 1.7 \pm 0.7 \pm 0.4) \times 10^{-2}$, 
$y_- =  (  2.1 \pm 2.2 \pm 0.5 \pm 1.1) \times 10^{-2}$,
$x_+  = (- 7.7 \pm 1.9 \pm 0.7 \pm 0.4) \times 10^{-2}$, and
$y_+  = (- 1.0 \pm 1.9 \pm 0.4 \pm 0.9) \times 10^{-2}$. 
 The first uncertainty is statistical, the second is systematic, and the third is due to the uncertainty on the strong-phase measurements.
These values are used to obtain 
$\gamma = \left(87\,^{+11}_{-12}\right)^\circ$, 
$r_B = 0.086^{+ 0.013}_{-0.014}$, and 
$\delta_B = (101 \pm 11)^\circ$,
 where $r_B$ is the ratio between the suppressed and favoured $B$-decay amplitudes and $\delta_B$
is the corresponding strong-interaction phase difference.
This measurement is combined with the result obtained using 2011 and 2012 data collected with the \lhcb experiment, to give
$\gamma = \left(80\,^{+10}_{\,-9}\right)^\circ$, 
$r_B = 0.080 \pm 0.011$, 
and $\delta_B = (110 \pm 10)^\circ$.
\end{abstract}

\vspace*{2.0cm}

\begin{center}
  Published in JHEP (2018) 2018: 176.
\end{center}

\vspace{\fill}

{\footnotesize 
\centerline{\copyright~\papercopyright. \href{\paperlicenceurl}{\paperlicence}.}}
\vspace*{1mm}

\end{titlepage}


\newpage
\setcounter{page}{2}
\mbox{~}
%
%
%
%

\cleardoublepage


\renewcommand{\thefootnote}{\arabic{footnote}}
\setcounter{footnote}{0}



\pagestyle{plain} 
\setcounter{page}{1}
\pagenumbering{arabic}


%

\section{Introduction}
\label{sec:Introduction}

The Standard Model (SM) description of \CP violation~\cite{Cabibbo:1963yz,Kobayashi:1973fv}
can be tested by overconstraining the angles of the Unitarity Triangle. 
The Cabibbo-Kobayashi-Maskawa (CKM) angle
$\gamma \equiv \mathrm{arg} (-V_{ud}^{\phantom{\ast}}V_{ub}^\ast/V_{cd}^{\phantom{\ast}}V_{cb}^\ast)$
is experimentally accessible through the interference between $\bar{b}\to\bar{c}u\bar{s}$ and $\bar{b}\to\bar{u}c\bar{s}$ transitions.
It is the only CKM angle easily accessible in tree-level processes and it can be measured 
with negligible uncertainty from theory~\cite{BrodZup}. 
Hence, in the absence of new physics effects at tree level, a precision measurement of $\gamma$ provides a SM benchmark that can be compared with other 
CKM-matrix observables more likely to be affected by physics beyond the SM. 
Such comparisons are currently limited by the uncertainty on direct measurements 
of $\gamma$, which is about $5\degrees$~\cite{HFLAV16} and is driven
by the \lhcb average.

The effects of interference between $\bar{b}\to\bar{c}u\bar{s}$ and $\bar{b}\to\bar{u}c\bar{s}$ transitions can be probed by studying \CP-violating observables in 
\BpmtoDKpm decays, where \D represents a \Dz or a \Dzb meson reconstructed in a final state that is common to both ~\cite{ads,GGSZ,BONDARGGSZ}.  
This decay mode has been studied at \lhcb with a wide range of $D$-meson final states to measure observables with 
sensitivity to $\gamma$~\cite{LHCb-PAPER-2013-068,LHCb-PAPER-2014-041,LHCb-PAPER-2015-014,LHCb-PAPER-2016-003}. 
In addition to these studies, other $B$ decays have also been used with a variety of techniques to determine 
$\gamma$~\cite{LHCb-PAPER-2014-028, LHCb-PAPER-2014-038, LHCb-PAPER-2015-020, LHCb-PAPER-2015-059}.

This paper presents a model-independent study of the 
decay mode \BpmtoDKpm, using \DtoKsPiPi and  \DtoKsKK decays (denoted \DtoKshh decays). The analysis utilises $pp$ collision data accumulated with \lhcb in 2015 and 2016 at a centre-of-mass energy of $\sqrt{s}=13$~\tev
and corresponding to a total integrated luminosity of $2.0\invfb$. 
The result is combined with the result obtained by \lhcb with the same analysis technique, using data collected in 2011 and 2012~(Run~1) at centre-of-mass energies of $\sqrt{s}=7$~\tev and $\sqrt{s}=8$~\tev~\cite{LHCb-PAPER-2014-041}. 

The sensitivity to $\gamma$ is obtained by comparing the distributions in the Dalitz plots of \DtoKshh 
decays from reconstructed $B^+$ and $B^-$ mesons~\cite{GGSZ,BONDARGGSZ}. For this comparison, the variation of the strong-phase difference between \Dz and \Dzb decay amplitudes within the Dalitz plot needs to be known.
An attractive, model-independent, approach makes use of direct measurements of the strong-phase 
variation over bins of the Dalitz plot~\cite{GGSZ,BPMODIND1,BPMODIND2}. The strong phase can be directly accessed by exploiting the quantum correlation of $\Dz\Dzb$ pairs from $\psi(3770)$ decays. Such measurements have been performed by the CLEO~collaboration~\cite{CLEOCISI} and have been used by the \lhcb~\cite{LHCb-PAPER-2014-041} and \belle~\cite{BELLEMODIND} collaborations to measure $\gamma$ in \BpmtoDKpm decays, and have also been used to study \BtoDKst decays ~\cite{BGGSZ,LHCb-PAPER-2016-006}. 
An alternative method relies on amplitude models of \DtoKshh decays, determined from flavour-tagged \DtoKshh decays, to predict the strong-phase variation over the Dalitz plot. This method has been used for a variety of $B$ decays~\cite{BABAR2005,BABAR2008,BABAR2010, BELLE2004,BELLE2006,BELLE2010,LHCb-PAPER-2014-017}.  

The separation of data into binned regions of the Dalitz plot leads to a loss of statistical sensitivity in comparison to using an amplitude model~\cite{BPMODIND1,BPMODIND2}. However, the advantage of using the direct strong-phase measurements resides in the model-independent nature of the systematic uncertainties. Where the direct strong-phase measurements are used, there is only a systematic uncertainty associated with the finite precision of such measurements. Conversely, systematic uncertainties associated with determining a phase from an amplitude model are difficult to evaluate, as common approaches to amplitude-model building break the optical theorem~\cite{Battaglieri:2014gca}. Therefore, the loss in statistical precision is compensated by reliability in the evaluation of the systematic uncertainty, which is increasingly important as the overall precision on the CKM angle $\gamma$ improves.

\section{Overview of the analysis}
\label{sec:principles}

The amplitude of the decay $B^- \to D K^-$, \DtoKshh can be written as a sum of the favoured 
$B^- \to D^0 K^-$ and  suppressed $B^- \to \Dzb K^-$ contributions as
\begin{equation}
A_B (m_-^2, m_+^2) \propto \, A_D(m_-^2,m_+^2) + r_B e^{i(\delta_B - \gamma)}A_{\Dbar}(m_-^2,m_+^2) ,
\label{eq:bamplitude}
\end{equation}
where  $m_-^2$ and $m_+^2$ are the squared invariant masses of the $\KS h^-$ and  $\KS h^+$ particle combinations, respectively,
that define the position of the decay in the Dalitz plot, $A_D(m_-^2,m_+^2)$ is the $\Dz \to \KS h^+ h^-$ decay amplitude, 
and $A_{\Dbar}(m_-^2,m_+^2)$ the  $\Dzb \to \KS h^+ h^-$ decay amplitude. 
The parameter $r_B$ is the ratio of the magnitudes of the $B^- \to \Dzb K^-$ 
and $B^- \to \Dz K^-$ amplitudes, while $\delta_B$ is their strong-phase difference.  
The equivalent expression for the charge-conjugated decay $B^+ \to D K^+$ is obtained by making 
the substitutions $\gamma \to -\gamma$ and $A_D(m_-^2,m_+^2) \leftrightarrow A_{\Dbar}(m_-^2,m_+^2)$.  
Neglecting \CP violation in charm decays,
the charge-conjugated amplitudes satisfy the relation $A_{\Dbar}(\msqmin,\msqplus) = A_D(\msqplus,\msqmin)$. 

The $D$-decay Dalitz plot is partitioned into $2\times N$ bins labelled from $i=-N$ to $i=+N$ (excluding zero), symmetric around  $\msqmin=\msqplus$  such that if $(\msqmin ,\msqplus)$ is in bin $i$ then $(\msqplus,\msqmin)$ is in bin $-i$. 
By convention, the positive values of $i$ correspond to bins for which $m_-^2 > m_+^2$.  
The strong-phase difference between the \Dz and \Dzb-decay amplitudes at a given point on the Dalitz plot is 
denoted as $\delta_D(m_-^2,m_+^2)$.  
The cosine of $\delta_D(m_-^2,m_+^2)$ weighted by the \D-decay amplitude and averaged 
over bin $i$ is written as $c_i$~\cite{GGSZ}, and is given by 
\begin{align}
c_i &\equiv \frac{\int_{i} d\msqmin \, d\msqplus \, |A_D(\msqmin,\msqplus)| |A_D(\msqplus,\msqmin)| \cos[\delta_D(\msqmin,\msqplus)-\delta_D(\msqplus,\msqmin)]}
{\sqrt{\int_{i} d\msqmin \, d\msqplus \, |A_D(\msqmin,\msqplus)|^2 \int_{i} d\msqmin \, d\msqplus \, |A_D(\msqplus,\msqmin)|^2}}\,,
\label{eq:ci}
\end{align}
where the integrals are evaluated over the phase space of bin $i$. An analogous expression can be written for $s_i$, which is 
the sine of the strong-phase difference weighted by the decay amplitude and averaged over the bin phase space.
The values of $c_i$ and $s_i$ have been directly measured by the CLEO collaboration, exploiting 
quantum-correlated $\Dz\Dzb$ pairs produced at the $\psi(3770)$ resonance~\cite{CLEOCISI}.

\begin{figure}[t]
\begin{center}
\includegraphics[width=0.48\textwidth]{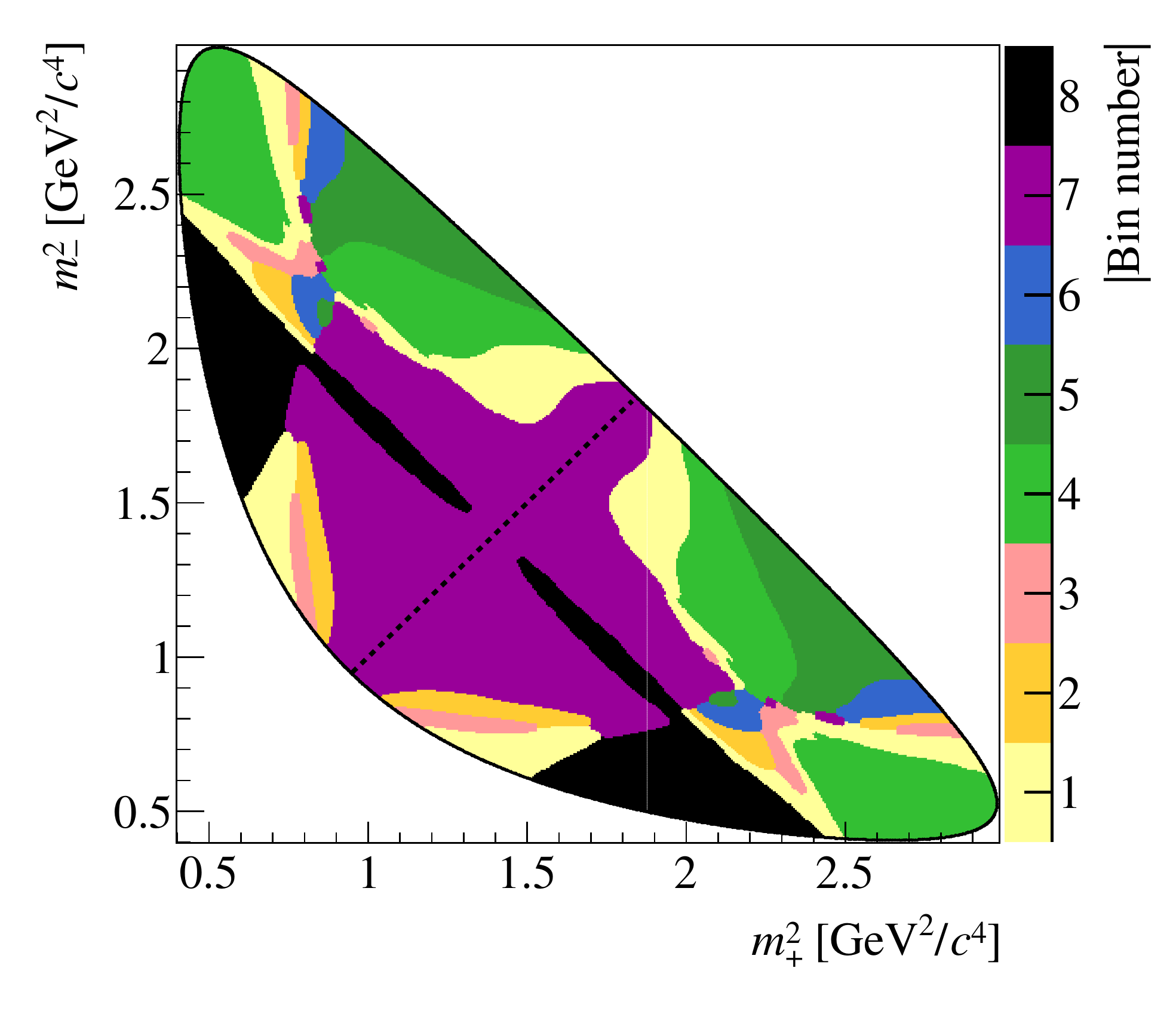}
\includegraphics[width=0.48\textwidth]{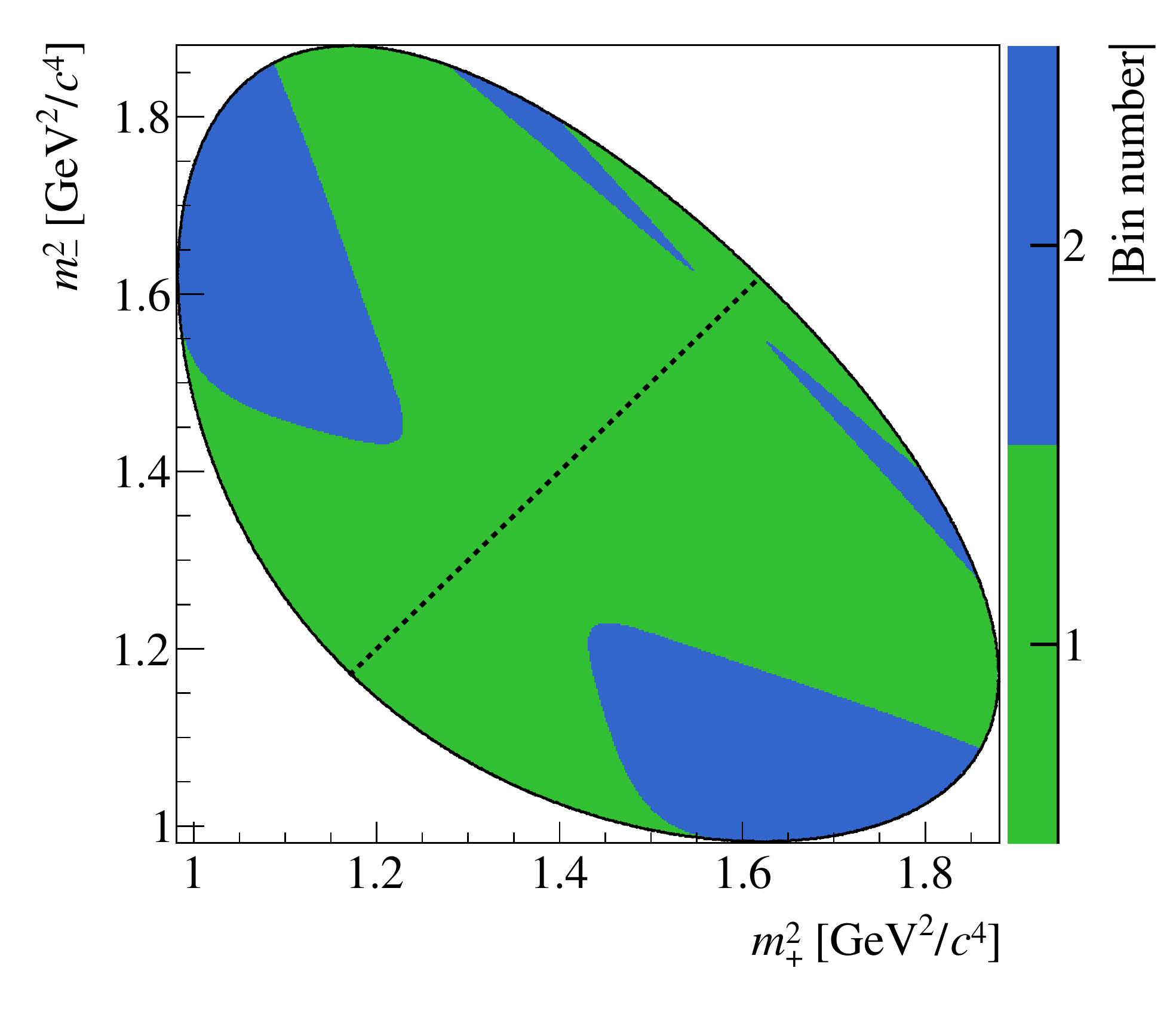}
\caption{\small Binning schemes for (left) \DtoKsPiPi decays and (right) \DtoKsKK decays. The diagonal line separates the positive and negative bins, where the positive bins are in the region in which $m^2_->m^2_+$ is satisfied. }
\label{fig:bins}
\end{center}
\end{figure}

The measurements of $c_i$ and $s_i$ are available in four different $2\times 8$ binning schemes 
for the \DtoKsPiPi decay. 
This analysis uses the `optimal binning' scheme where the bins have been 
chosen to optimise the statistical sensitivity to $\gamma$, as described in Ref.~\cite{CLEOCISI}. The optimisation was performed assuming a strong-phase 
difference distribution as predicted by the \babar model presented in Ref.~\cite{BABAR2008}. 
For the \KsKK final state, three choices of binning schemes are available, containing $2\times2$, $2\times3$, and $2\times4$ bins.  The guiding model used to determine the bin boundaries
is taken from the \babar study described in Ref.~\cite{BABAR2010}. 
The $2\times 2$ binning scheme is chosen, due to the low signal yields in the \DtoKsKK mode.
The same choice of bins was used in the LHCb Run~1 analysis \cite{LHCb-PAPER-2014-041}.
The measurements of $c_i$ and $s_i$ are not biased by the use of a specific amplitude model in defining 
the bin boundaries. The choice of the model only affects this analysis to the extent that a poor model description of the underlying 
decay would result in a reduced statistical sensitivity of the $\gamma$ measurement. The binning choices for 
the two decay modes are shown in Fig.~\ref{fig:bins}.

The physics parameters of interest, $r_B$, $\delta_B$, and $\gamma$, are translated into four \CP obser-vables~\cite{BABAR2005} that 
are measured in this analysis. These observables are defined as
\begin{equation}
x_\pm \equiv r_B \cos (\delta_B \pm \gamma) {\rm\ \ and\ } \; y_\pm \equiv r_B \sin (\delta_B \pm \gamma).
\label{eq:xydefinitions}
\end{equation}
It follows from Eq.~\eqref{eq:bamplitude} that the expected numbers of \Bp and \Bm decays in bin $i$, $N^{+}_i$ and $N^-_i$, are given by 
\begin{align}
\begin{split}
N_{\pm i}^+ &= h_{B^+} \left[ F_{\mp i} + (x_+^2 + y_+^2) F_{\pm i} + 2 \sqrt{F_i F_{-i}} ( x_+ c_{\pm i} - y_+ s_{\pm i}) \right], \\
N_{\pm i}^- &= h_{B^-} \left[ F_{\pm i} + (x_-^2 + y_-^2) F_{\mp i} + 2 \sqrt{F_i F_{-i}} ( x_- c_{\pm i} + y_- s_{\pm i}) \right],
\label{eq:populations}
\end{split}
\end{align}
where $F_i$ are the fractions of decays in bin $i$ of the $\Dz\to\Kshh$ Dalitz plot,
and $h_{B^\pm}$ are normalisation factors, which can be different for $B^+$ and $B^-$ due to production, detection, and \CP asymmetries. In this measurement, the integrated yields are not used to provide information on \xpm and \ypm, and so the analysis is insensitive to such effects.
From Eq.~\eqref{eq:populations} it is seen that studying the distribution of candidates over the \DtoKshh Dalitz plot gives access to the 
$x_\pm$ and $y_\pm$ observables. 
The detector and selection requirements placed on the data lead to a non-uniform efficiency over the Dalitz plot, which
affects the $F_i$ parameters.
The efficiency profile for the signal candidates is denoted as $\eta(\msqmin,\msqplus)$.
The parameters $F_i$ can then be expressed as
\begin{equation}
\label{eq:fi}
F_i = \frac{\int_{i} d\msqmin d\msqplus |A_{D}(\msqmin,\msqplus)|^2 \, \eta(\msqmin,\msqplus) }{\sum_j \int_{j} d\msqmin d\msqplus |A_{D}(\msqmin,\msqplus)|^2\, \eta(\msqmin,\msqplus) }\,.
\end{equation}
The values of $F_i$ are determined from the control decay mode \BztoDstmuX, where the \Dstarm meson decays 
to \Dzb\pim and the \Dzb meson decays to either the \KsPiPi or \KsKK final state. 
The symbol $X$
indicates other particles which may be produced in the decay but are not reconstructed. 
Samples of simulated events are used 
to correct for the small differences in efficiency arising through unavoidable differences in selecting \BztoDstmuX 
and \BtoDK decays, as discussed further in Sect.~\ref{sec:dmu}.

In addition to \BpmtoDKpm and \BztoDstmuX candidates, \BpmtoDpipm decays are selected. These 
provide an important control sample that is used to constrain the invariant-mass shape of the \BpmtoDKpm signal,
as well as to determine the yield of \BpmtoDpipm decays misidentified as \BpmtoDKpm candidates. 
Note that this channel is not optimal for determining the values of $F_i$
as the small level of \CP violation in the decay leads to a significant systematic uncertainty,
as was reported in Ref.~\cite{LHCb-PAPER-2012-027}. 
This uncertainty is eliminated when using the flavour-specific semileptonic decay,
in favour of a smaller systematic uncertainty associated with efficiency differences. 

The effect of \Dz--\Dzb mixing was ignored in the above discussion. If the parameters $F_i$ are obtained from \BztoDstmuX, where the \Dstarm decays 
to \Dzb\pim, \Dz--\Dzb mixing has been shown to lead to a bias of approximately $0.2^\circ$ 
in the $\gamma$ determination~\cite{BPV}, which is negligible for the current analysis. The effects of \CP violation 
in the neutral kaon system and of the different nuclear interaction cross-sections for \Kz 
and \Kzb mesons are discussed in Sect.~\ref{sec:syst}, where a systematic uncertainty is assigned.

\section{Detector and simulation}
\label{sec:Detector}

The \lhcb detector~\cite{Alves:2008zz,LHCb-DP-2014-002} is a single-arm forward
spectrometer covering the \mbox{pseudorapidity} range $2<\eta <5$,
designed for the study of particles containing \bquark or \cquark
quarks. The detector includes a high-precision tracking system
consisting of a silicon-strip vertex detector surrounding the $pp$
interaction region, a large-area silicon-strip detector located
upstream of a dipole magnet with a bending power of about
$4{\mathrm{\,Tm}}$, and three stations of silicon-strip detectors and straw
drift tubes placed downstream of the magnet.
The polarity of the dipole magnet is reversed periodically throughout data-taking.
The tracking system provides a measurement of momentum, \ptot, of charged particles with
 relative uncertainty that varies from 0.5\% at low momentum to 1.0\% at 200\gevc.
The minimum distance of a track to a primary vertex (PV), the impact parameter (IP), 
is measured with a resolution of $(15+29/\pt)\mum$,
where \pt is the component of the momentum transverse to the beam, in\,\gevc.
Different types of charged hadrons are distinguished using information
from two ring-imaging Cherenkov detectors. 
Photons, electrons, and hadrons are identified by a calorimeter system consisting of
scintillating-pad and preshower detectors, an electromagnetic
calorimeter and a hadronic calorimeter. Muons are identified by a
system composed of alternating layers of iron and multiwire
proportional chambers.

The online event selection is performed by a trigger, 
which consists of a hardware stage based on information from the calorimeter and muon
systems, followed by a software stage, which applies a full event
reconstruction.
At the hardware trigger stage, events are required to have a muon with high \pt or a
hadron, photon or electron with high transverse energy in the calorimeters. For hadrons,
the transverse energy threshold is 3.5\gev.
The software trigger requires a two-, three- or four-track
secondary vertex with a significant displacement from any primary
$pp$ interaction vertex. At least one charged particle
must have transverse momentum $\pt > 1.6\gevc$ and be
inconsistent with originating from a PV.
A multivariate algorithm~\cite{BBDT} is used for
the identification of secondary vertices consistent with the decay
of a \bquark hadron. Small changes in the trigger thresholds were made throughout both years of data taking.

In the simulation, $pp$ collisions are generated using
\pythia~8~\cite{Sjostrand:2007gs, *Sjostrand:2006za} 
 with a specific \lhcb configuration~\cite{LHCb-PROC-2010-056}.  Decays of hadronic particles
are described by \evtgen~\cite{Lange:2001uf}, in which final-state
radiation is generated using \photos~\cite{Golonka:2005pn}. The
interaction of the generated particles with the detector, and its response,
are implemented using the \geant
toolkit~\cite{Allison:2006ve, *Agostinelli:2002hh} as described in
Ref.~\cite{LHCb-PROC-2011-006}.

\section{Event selection and fit to the invariant-mass spectrum for
\texorpdfstring{$\boldsymbol{\BpmtoDKpm}$}{B to DK} and \texorpdfstring{$\boldsymbol{\BpmtoDpipm}$ decays}{B to Dpi}
}
\label{sec:selection}

Decays of \KS mesons to the $\pip\pim$ final state are reconstructed in two categories,
the first containing \KS mesons that decay early enough for the pions to be reconstructed in the vertex detector and the
second containing \KS mesons that decay later such that track segments of the pions cannot be formed in the vertex detector. These categories are
referred to as \emph{long} and \emph{downstream}, respectively. The candidates in the long category have better mass, momentum and vertex resolution 
than those in the downstream category. Hereinafter, \Bpm candidates are denoted long or downstream depending on which category of \KS candidate they contain.

For many of the quantities used in the selection and analysis of the data,
a kinematic fit~\cite{Hulsbergen:2005pu} is imposed on the full \Bpm decay chain. 
Depending on the quantity being calculated, 
the \D and \KS candidates may be constrained to have their known masses~\cite{PDG2017}, as described below. 
The fit also constrains the \Bpm candidate momentum vector to point towards the associated PV, defined as the PV for which the candidate has the smallest IP significance.
These constraints improve the resolution of the calculated quantities, and thus help improve separation
between signal and background decays.
Furthermore, it improves the resolution on the Dalitz plot coordinates and ensures that all candidates 
lie within the kinematically allowed \DtoKshh phase space.

The $D$ ($\KS$) candidates are required to be within $25\mevcc$ 
($15\mevcc$) of their known 
mass~\cite{PDG2017}. These requirements are placed on masses obtained using
kinematic fits in which all constraints are applied except for that on the mass under consideration.
Combinatorial background is primarily suppressed through the use of a boosted decision tree (BDT) multivariate classifier~\cite{Breiman,AdaBoost}. 
The BDT is trained on simulated signal events and background taken from the high \Bpm mass sideband (5800--7000\mevcc). Separate BDTs are trained for the long and downstream categories.

Each BDT uses the same set of variables: 
the \chisq of the kinematic fit of the whole decay chain;
\ptot and \pt of the companion, \D, and \Bpm after the kinematic refit (here and in the following, companion refers to the final state \pipm or \Kpm meson produced in the $\Bpm\to\D\hadron^\pm$ decay); 
the vertex quality of the \KS, \D, and \Bpm candidates;
the distance of closest approach between tracks forming the \D and \Bpm vertices;
the cosine of the angle between the momentum vector and the vector between the production and decay vertices
of a given particle, for each of the \KS, \D, and \Bpm candidates;
the minimum and maximum values of the \chisqip of the pions from both the \D and \KS decays, where \chisqip is defined as the difference in \chisq of the PV fit with and without the considered 
 particle;
the \chisqip for the companion, \KS, \D, and \Bpm candidates; 
the \Bpm flight-distance significance; 
the radial distance from the beamline to the \D and \Bpm-candidate vertices; 
and a \Bpm \emph{isolation variable}, which is designed to ensure the \Bpm candidate is well isolated from other 
tracks in the event. The \Bpm isolation variable is the asymmetry between the \pt of the signal candidate and 
the sum of the \pt of other tracks in the event that lie within a distance of 1.5 
in $\eta$--$\phi$ space, where $\phi$ is the azimuthal angle measured in radians. 
Candidates in the data samples that have a BDT output value below a threshold are rejected.
An optimal threshold value is determined for each of the two BDTs, using a series of pseudoexperiments to obtain the 
values that provide the best sensitivity to \xpm and \ypm.  Across all \BtoDK channels this requirement is found to reject 99.1\,\% of the combinatorial background in the high \B mass sideband that survives all other requirements, while having an efficiency of 92.4\,\% on simulated \BtoDK signal samples.

Particle identification (PID) requirements are placed on the companion to separate \BpmtoDKpm and \BpmtoDpipm candidates, and on the charged decay products of the $D$ meson to remove cross-feed between different $D \to \Kshh$ decays. 
To ensure good control of the PID performance it is required that information from the RICH detectors is present.
To remove background from $D\to\pip\pim\pip\pim$ or 
$D\to\pip\pim\Kp\Km$ decays, long \KS candidates are required to have travelled a significant distance from 
the \D vertex. This requirement is not necessary for downstream candidates. Similarly, the \D decay vertex is required be significantly displaced from the 
\Bpm decay vertex in order to remove charmless \Bpm decays.

\begin{figure}[t]
\centering
\includegraphics[width=0.45\textwidth]{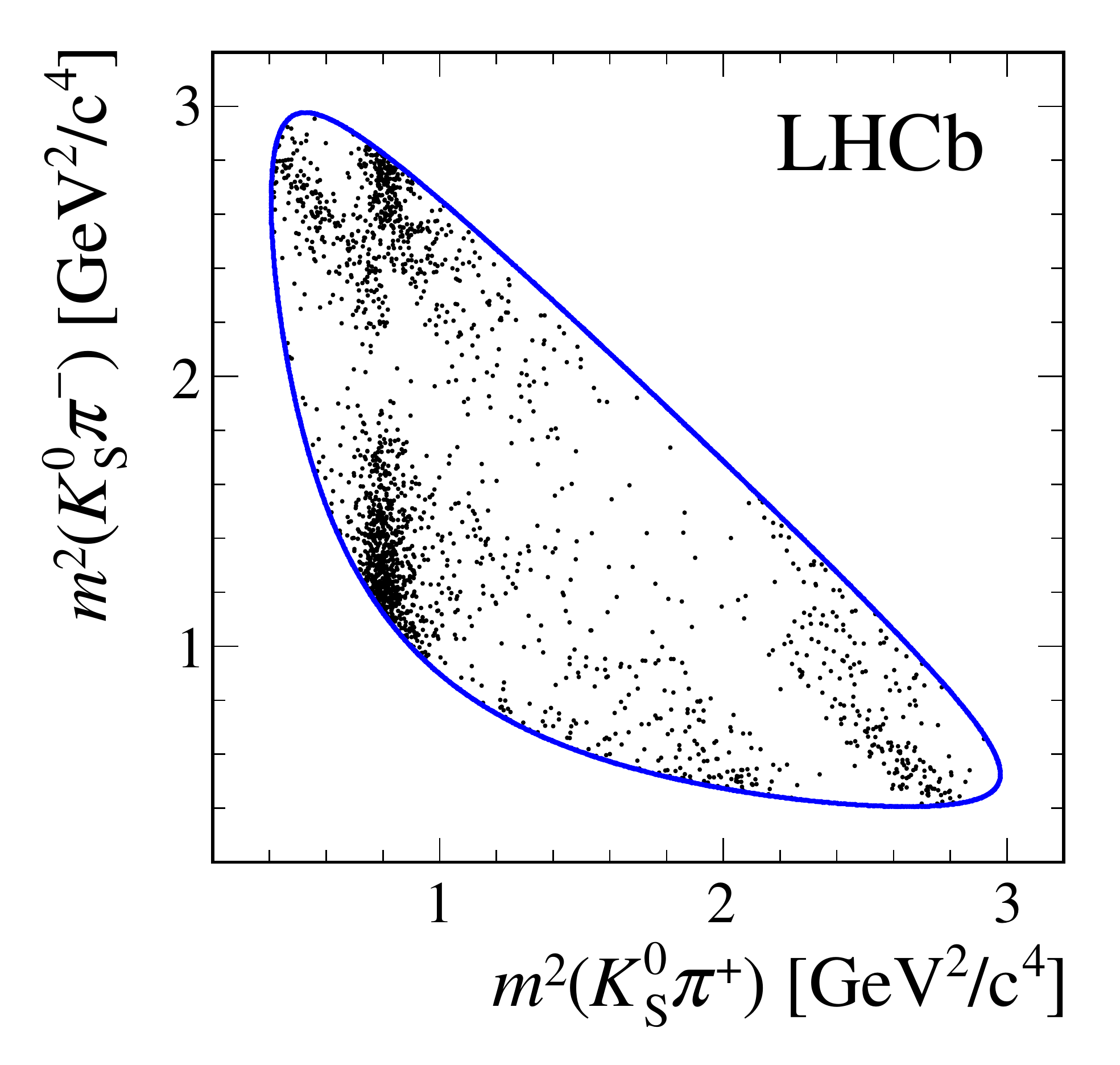}
\includegraphics[width=0.45\textwidth]{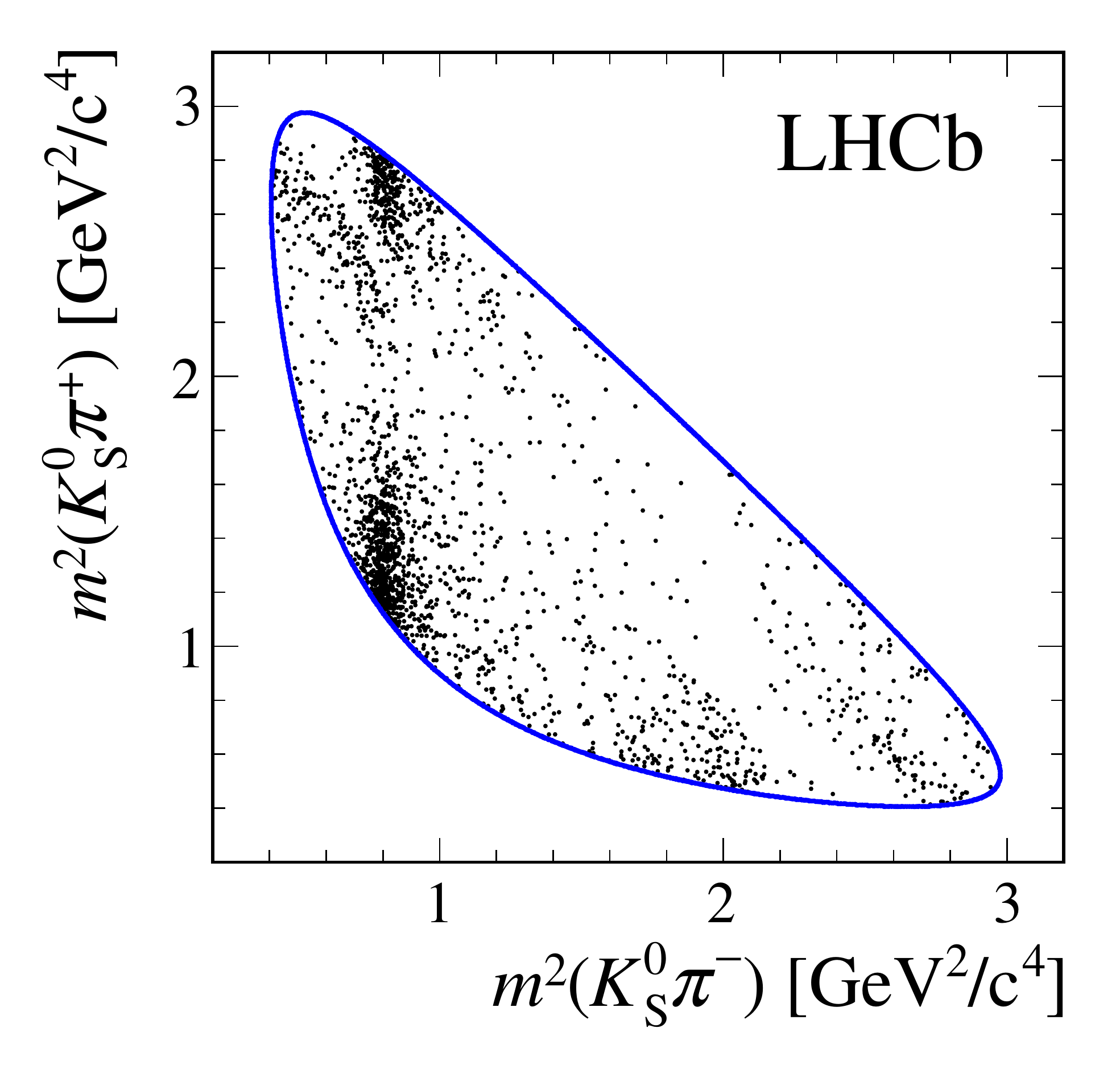}\\
\includegraphics[width=0.45\textwidth]{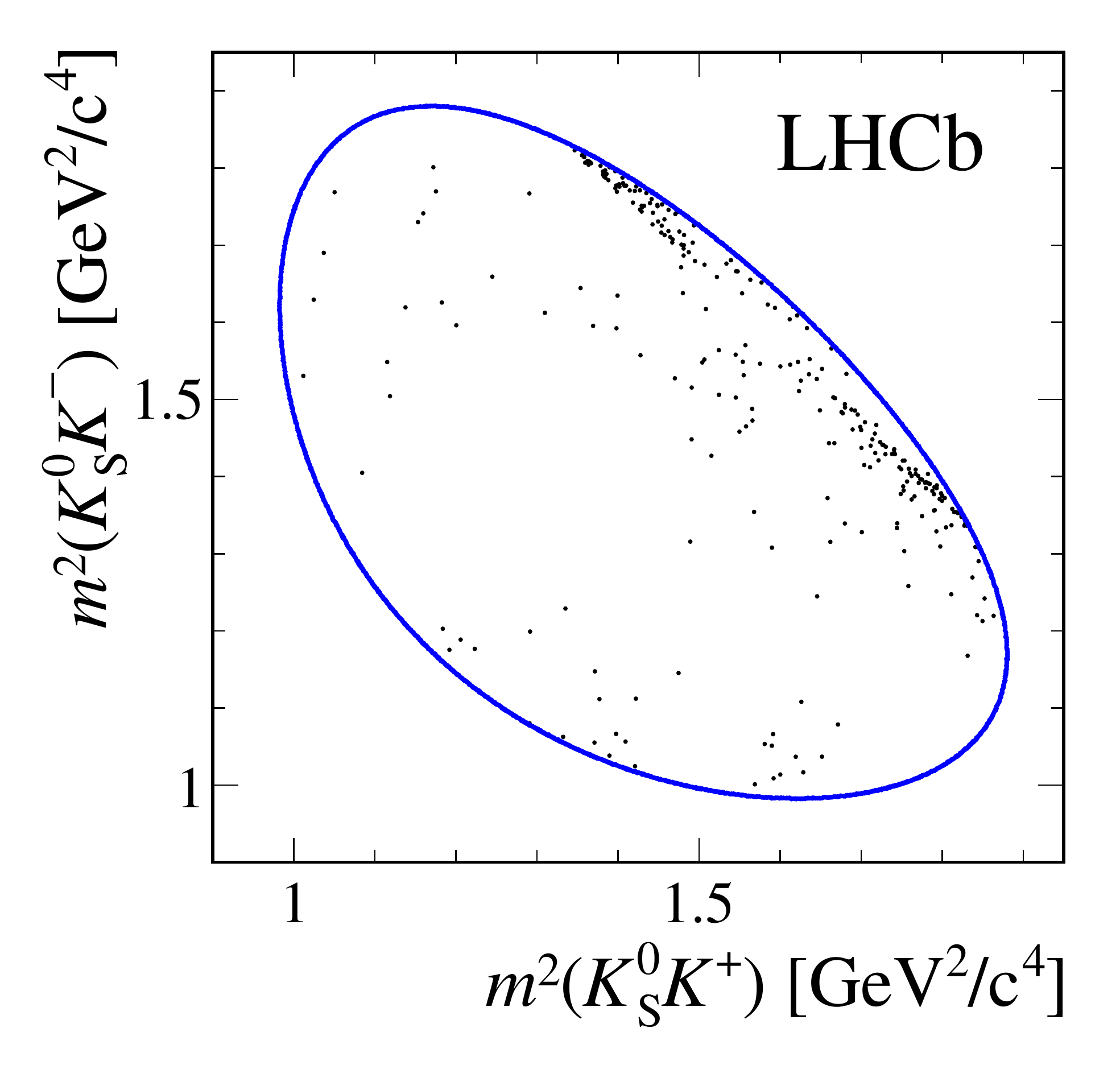}
\includegraphics[width=0.45\textwidth]{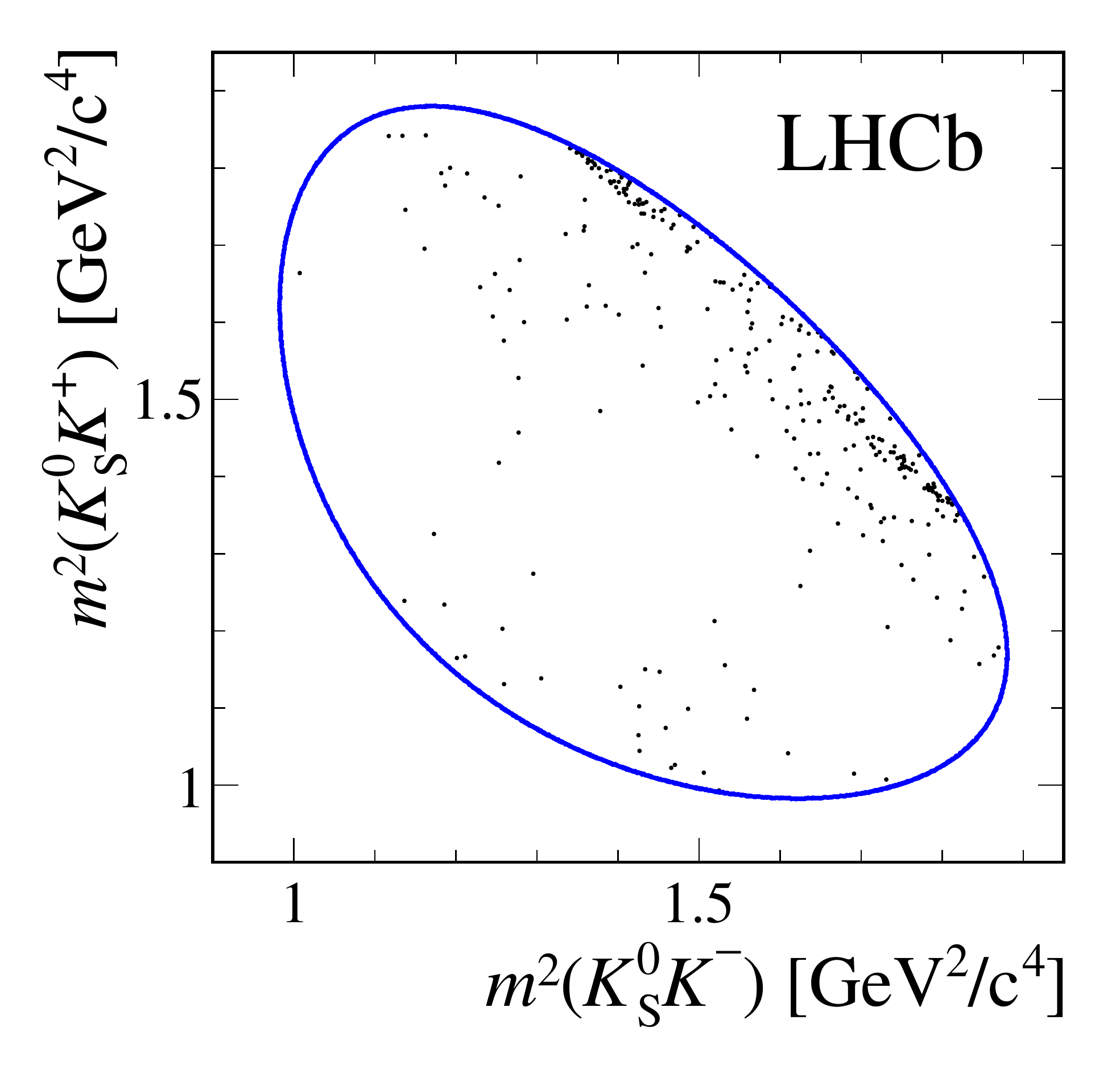}
\caption{\small Dalitz plots of long and downstream (left) $\Bp\to D \Kp$ and (right) $\Bm\to D \Km$ candidates  for (top) \DtoKsPiPi and (bottom) \DtoKsKK decays in which the reconstructed invariant mass of the \Bpm candidate is in a region of $\pm 25\mevcc$ around the \Bpm mass. The narrow region is chosen to obtain high purity, as no background subtraction has been made. The Dalitz coordinates 
are calculated using the results of a kinematic fit in which the \D and \KS masses are constrained to their known values. The blue lines show the kinematic boundaries of the decays.}
\label{fig:dalitz}
\end{figure}

The Dalitz plots for \BpmtoDKpm candidates in a narrow region of $\pm 25\mevcc$ around the \Bpm mass are shown in 
Fig.~\ref{fig:dalitz}, for both \DtoKshh final states samples.  Separate plots are shown for $B^+$ and $B^-$ decays. The Dalitz coordinates 
are calculated from the kinematic fit with all mass constraints applied.

\begin{figure}[t]
\centering
\includegraphics[width=0.98\textwidth]{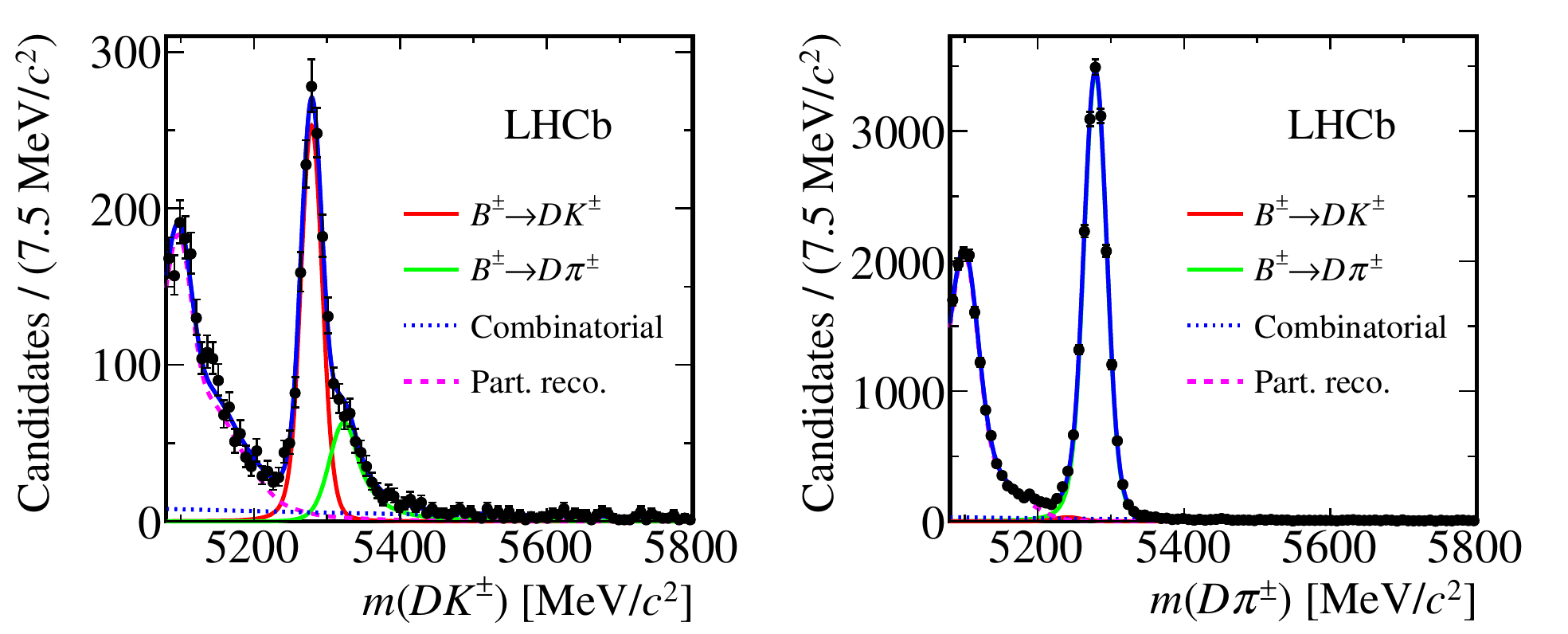} \\
\includegraphics[width=0.98\textwidth]{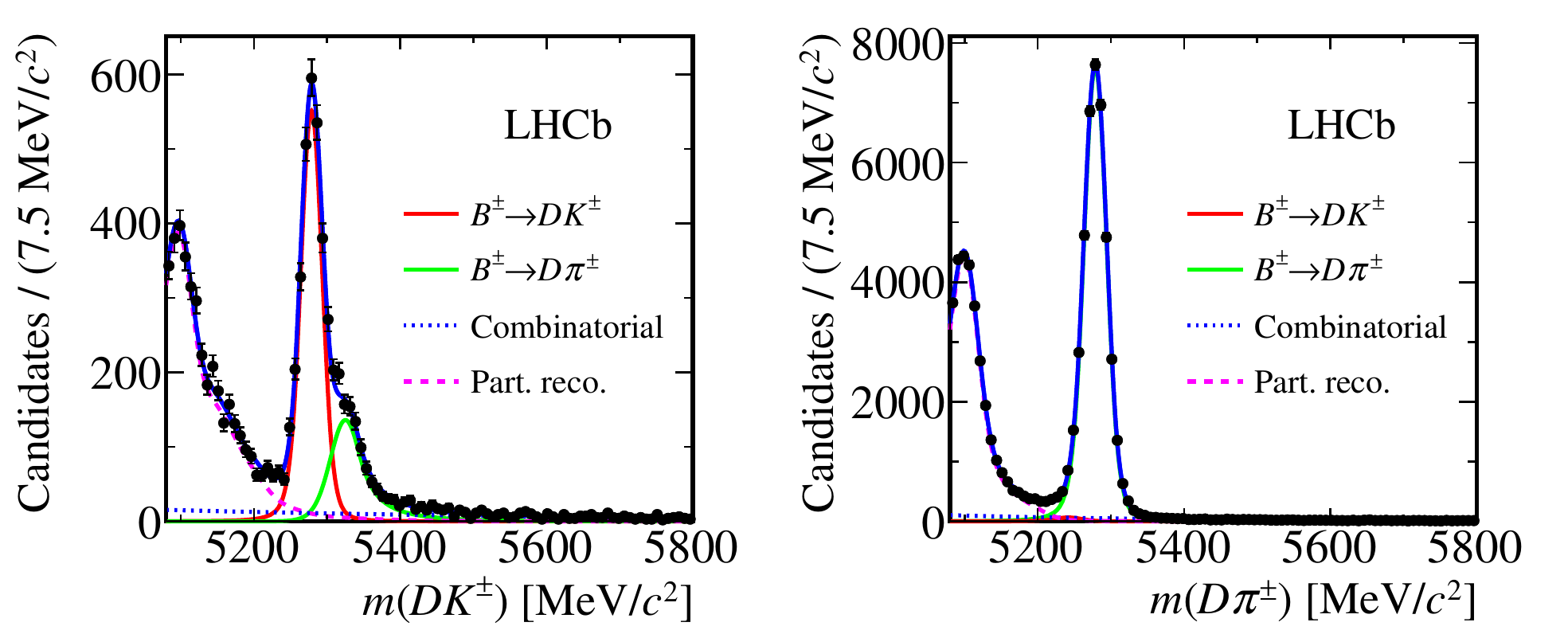}
\caption{\small Invariant-mass distributions of (left) \BpmtoDKpm and  (right) \BpmtoDpipm candidates, with \DtoKsPiPi, shown separately for the (top) long  and (bottom) downstream \KS categories. Fit results, including the signal component and background components due to misidentified companions, partially reconstructed decays and combinatorial background, are also shown.}
\label{fig:mass_kspipi}
\end{figure}

\begin{figure}[t]
\centering
\includegraphics[width=0.98\textwidth]{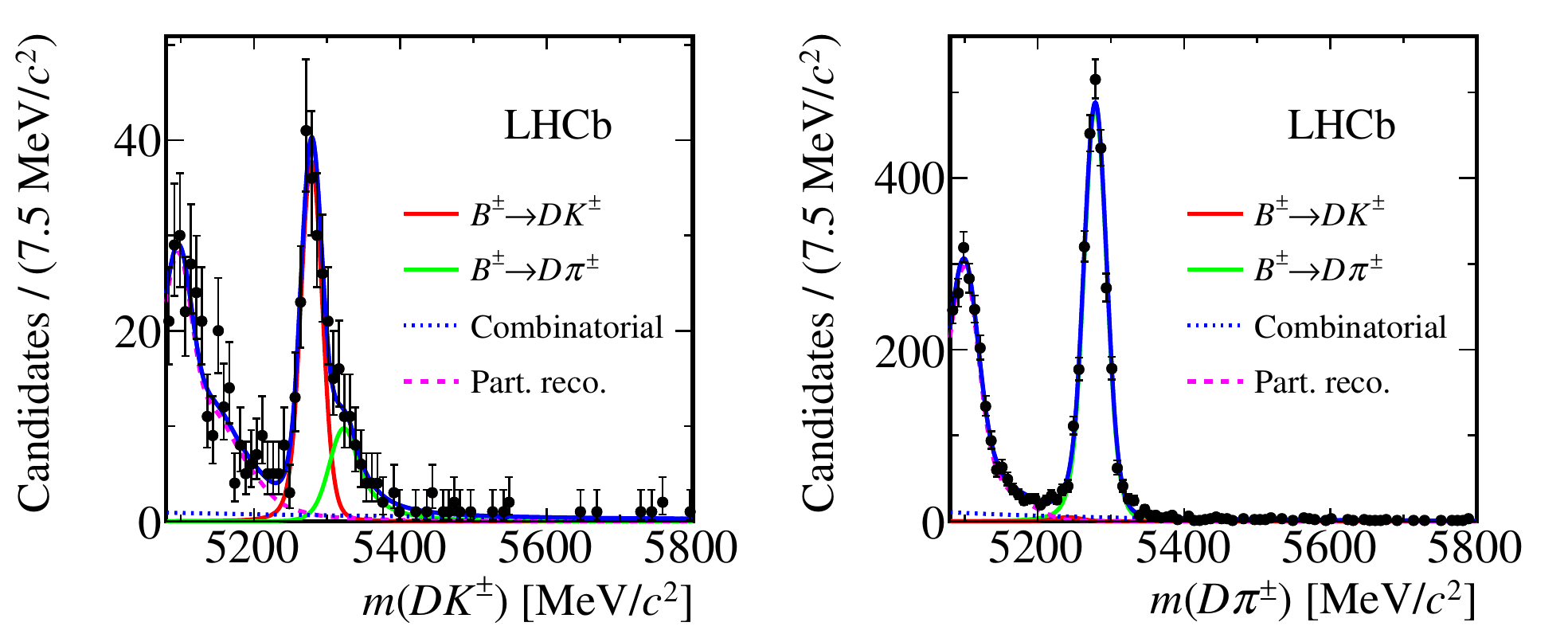} \\
\includegraphics[width=0.98\textwidth]{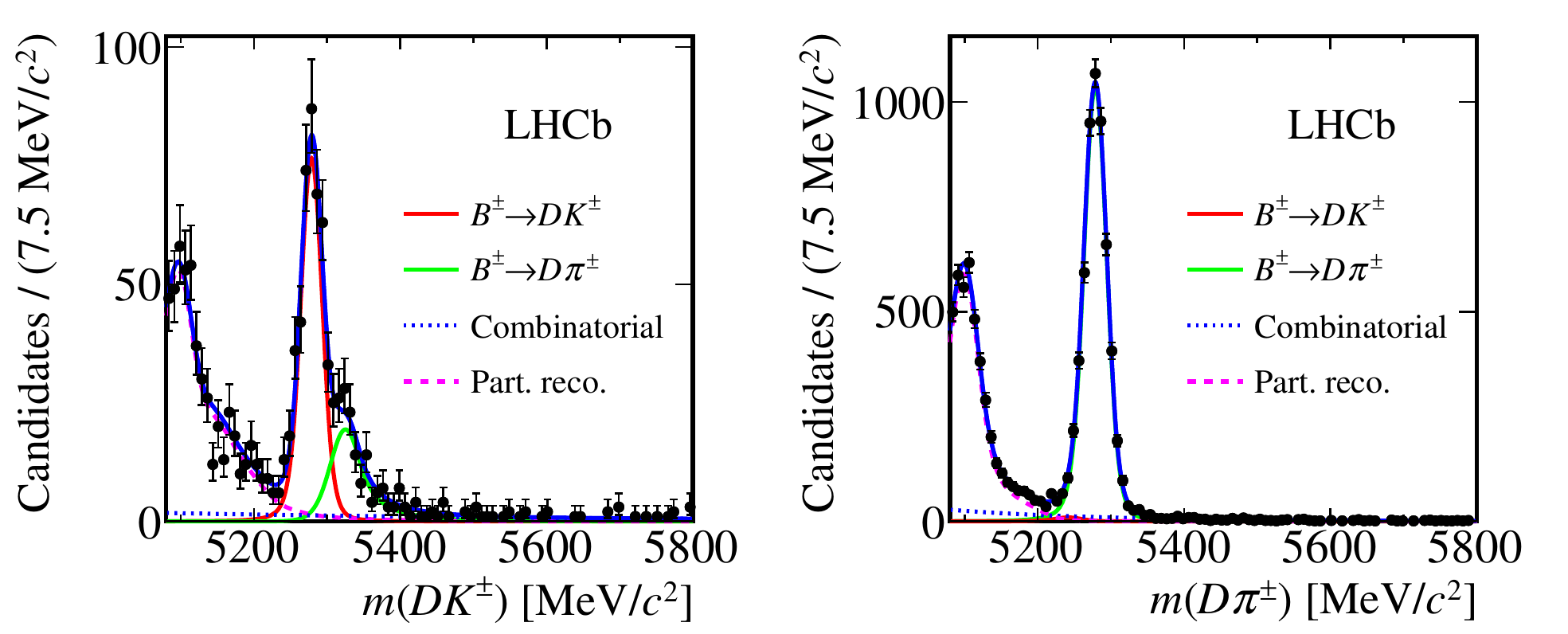}
\caption{\small Invariant-mass distributions of (left) \BpmtoDKpm and  (right) \BpmtoDpipm candidates, with \DtoKsKK, shown separately for the (top) long  and (bottom) downstream \KS categories. Fit results, including the signal component and background components due to misidentified companions, partially reconstructed decays and combinatorial background, are also shown.}
\label{fig:mass_kskk}
\end{figure}

In order to determine the parameterisation of the signal and background components
that are used in the fit of partitioned regions of the Dalitz plot described in Sect.~\ref{sec:dpfit}, an extended maximum likelihood fit to the invariant-mass distributions of the \Bpm candidates is performed, 
in which the \Bp and \Bm candidates in all of the Dalitz bins are combined. The invariant mass of each \Bpm candidate is calculated using the results of a kinematic fit in which the \D and \KS masses are constrained to their known values. The sample is split into \BtoDK and \BtoDpi candidates, by \D decay mode and by \KS category. In order to allow sharing of some parameters,
the fit is performed simultaneously for all of the above categories. The projections of the fit and the invariant-mass distributions of the selected \Bpm candidates are shown in Figs.~\ref{fig:mass_kspipi} and~\ref{fig:mass_kskk} for \DtoKsPiPi and \DtoKsKK candidates, respectively. 
The fit range is between 5080\mevcc and 5800\mevcc in the \Bpm candidate invariant mass.


The peaks corresponding to actual \BtoDK and \BtoDpi candidates are fitted with a sum of two Crystal Ball~\cite{Skwarnicki:1986xj} functions, 
which are parameterised as
\begin{align}\label{eq:CB_definition}
  \text{CB}(m,\mu,\sigma,\alpha, n)\propto \left \lbrace
  \begin{array}{ll}
    \exp \left [ -\frac{1}{2}\left(\frac{m-\mu}{\sigma}\right)^2 \right] & \text{if   } (m-\mu)/\sigma > -\alpha  \\
    A\left(B-\frac{m-\mu}{\sigma}\right)^{-n} & \text{otherwise,}\end{array}\right.
\end{align}
where $\alpha >0$, and
\begin{align}
  A &= \left(\frac{n}{\alpha}\right)^n\exp [-\alpha^2/2]\,, \\
  B &= \frac{n}{\alpha}-\alpha\,.
\end{align}The sum is implemented such that the Crystal Ball functions have tails pointing in either direction. They share a common width, $\sigma$, and mean, $\mu$. In practice, the signal probability density function (PDF) is defined as
\begin{align}
  f_{\text{signal}}(m,\mu,\sigma,\alpha_\text{L}&, n_\text{L}, \alpha_\text{R}, n_\text{R}, f_{\text{CB}}) \notag
  \\ &= f_{\text{CB}}\cdot \text{CB}(m, \mu, \sigma, \alpha_\text{L}, n_\text{L}) + (1-f_{\text{CB}})\cdot \text{CB}(m,\mu,-\sigma,\alpha_\text{R},n_\text{R}).
\end{align} 
The tail parameters, $n_{\text{L},\text{R}}$ and $\alpha_{\text{L},\text{R}}$, are fixed from simulation,
while the other parameters are left as free parameters in the fit. Separate tail parameters, $f_\text{CB}$, and
$\sigma$ are used for long and downstream candidates. Different widths are used for the \BpmtoDKpm 
and \BpmtoDpipm channels, with their ratio $r_{\sigma} = \sigma_{DK}/\sigma_{D\pi}$ shared between 
all categories. The mean is shared among all categories. 
The yield of \BpmtoDpipm decays in each \KS and \D-meson decay category, $N^\text{cat}(\BpmtoDpipm)$, is determined in the fit. Instead of 
fitting the yield of \BpmtoDKpm decays directly in each category, it is determined from the $\BpmtoDpipm$ yield in the corresponding category and the ratio
\begin{align}
\mathcal{R} \equiv \frac{N^\text{cat}(\BpmtoDKpm)}{\eps^\text{cat}_\text{PID}(\BpmtoDKpm)}\bigg/\frac{N^\text{cat}(\BpmtoDpipm)}{\eps^\text{cat}_\text{PID}(\BpmtoDpipm)} \, ,
\end{align}
which is a free parameter in the fit. The category-dependent PID efficiencies, \linebreak $\eps^\text{cat}_\text{PID}(\Bpm\to\D h^\pm)$, are taken into account, so that a single $\mathcal{R}$ parameter can be shared between all categories in the fit. How these efficiencies are obtained is described below.
As the parameter $\mathcal{R}$ is efficiency corrected, it is equal to the ratio of branching fractions between the $\BpmtoDKpm$ and $\BpmtoDpipm$ decay modes. The measured ratio is found to be ${\mathcal{R}}=(7.66\pm 0.14)\,\%$, where the uncertainty is statistical only, and this is consistent with the expected value of $(7.8\pm0.4)\,\%$~\cite{PDG2017}.

The background consists of random track combinations, 
partially reconstructed \PB decays, and \BpmtoDhpm decays in which the companion has been misidentified.  
The random track combinations are modelled by an exponential PDF. The slopes of the exponentials are free parameters in the fit to the data. These slopes are 
independent for each of the \BpmtoDpipm categories, while they are shared for the \BpmtoDKpm categories to improve the stability of the fit. When these slopes are allowed to be independent, the fit returns results that are statistically compatible.

In the \BpmtoDKpm sample there is a clear contribution from \BpmtoDpipm decays in which 
the companion particle is misidentified as a kaon by the RICH system.
The rate for \BpmtoDKpm decays to be misidentified and placed in the \BpmtoDpipm
sample is much lower due to the reduced branching fraction.
Nevertheless, this contribution is still accounted for in the fit.
The yields of these backgrounds are fixed in the fit, using knowledge of misidentification 
efficiencies and the fitted yields of reconstructed decays
with the correct particle hypothesis. The misidentification efficiencies are obtained from large samples 
of $\Dstarpm \to \DorDbar^0 \pi^\pm$, $\DorDbar^0 \to K^\mp \pi^\pm$ decays.
These decays are selected using only kinematic variables in order to provide
pure samples of $\Kmp$ and $\pipm$ that are unbiased in the PID variables. 
The PID efficiency is parameterised as a function
of the companion momentum and pseudorapidity, and the charged-particle multiplicity in
the event. The calibration sample is weighted so that the distribution of these variables matches that of the candidates in the signal region of the
 \Bpm sample, thereby 
ensuring that the measured PID performance is representative for the samples used in this measurement.  The efficiency to identify 
a kaon correctly is found to be approximately 86\,\%, while that for a pion is approximately 97\,\%.  
The PDFs of the backgrounds due to misidentified companion particles are determined using data. 
As an example, consider the case of true \BpmtoDpipm decays misidentified as \BpmtoDKpm candidates.
The \sPlot{} method~\cite{Pivk:2004ty} is used on the \BpmtoDpipm sample in order to isolate the
contribution from the signal decays. The \Bpm invariant mass is then calculated using the kaon mass hypothesis for the companion
pion, and weighting by PID efficiencies in order to properly reproduce
the kinematic properties of pions misidentified as kaons in the signal \BpmtoDKpm sample. The weighted distribution is fitted 
with a sum of two Crystal Ball shapes. The fitted parameters are 
subsequently fixed in the fit to the \Bpm invariant-mass spectrum, with the procedure applied separately
for long and downstream candidates. An analogous approach is used to determine the shape of the misidentified
\BpmtoDKpm contribution in the \BpmtoDpipm sample.

Partially reconstructed $b$-hadron decays contaminate the sample predominantly at invariant masses smaller than that of the signal peak. 
These decays contain an unreconstructed  pion or photon, which predominantly comes from an intermediate resonance. 
There are contributions from $B^0 \to D^{*\pm} h^\mp$ and $B^\pm \to D^{*0} h^\pm$ decays in all channels (denoted as $\B\to\D^*h^\pm$ decays),
while $B^\pm \to \D \rho^\pm$ and $B^\pm \to \D \Kstarpm$ decays contribute to the
\BpmtoDpipm and \BpmtoDKpm channels, respectively. In the \BtoDK channels there is also a contribution from $\Bs\to\Dzb\pi^+\Km$ ($\Bsb\to\Dz\pi^-\Kp$) decays where the charged pion is not reconstructed. 
The invariant-mass distributions of these backgrounds depend 
on the spin and mass of the missing particle, as described in Ref.~\cite{LHCb-PAPER-2017-021}. The shape of the background from \Bs decays is based on the results of Ref.~\cite{LHCb-PAPER-2014-036}.
Additionally, each of the above backgrounds of \BpmtoDpipm decays can contribute in the \BpmtoDKpm channels if the pion is misidentified.
The inverse contribution is negligible and is neglected.
The shapes for the decays in which a pion is misidentified as a kaon are 
parameterised with semi-empirical PDFs formed from sums of Gaussian and error functions. 
The parameters of these backgrounds are fixed to the results of fits to data from two-body $D$ decays~\cite{LHCb-PAPER-2017-021}, 
where they were obtained with a much larger data sample.
However, the width of the resolution function and a shift along the \Bpm mass
are allowed to differ in order to accommodate small differences between the \D decay modes. 

In each of the \BpmtoDpipm channels, the total yield of the partially reconstructed background is fitted independently. The relative amount of each $\B \to \D^* \pi^\mp$ mode is fixed from efficiencies obtained from simulation and known branching fractions, while the fraction of $B^\pm \to \D \rho^\pm$ decays is left free. 
In the \BpmtoDKpm channels, the yield of the \linebreak $\Bs\to\Dzb\pi^+\Km$ background is fixed relative to the corresponding \BtoDpi yield, using efficiencies from simulation and the known branching fraction.
The total yield of the remaining partially reconstructed backgrounds is expressed via a single fraction,
$\mathcal{R}_{\D\kaon/\D\pi}^\text{low}$,
relative the \BpmtoDpipm yields. It is free in the fit, and common to all channels after taking into account the different particle-identification efficiencies.
The relative amount of each $\B \to \D^* \kaon^\mp$ mode is fixed using efficiencies from simulation and known branching fractions, while the fraction of $B^\pm \to \D \Kstar^\pm$ decays is fixed using the results of Ref.~\cite{LHCb-PAPER-2017-021}. The yields of the partially reconstructed modes with a companion pion misidentified as a kaon are fixed via the known PID efficiencies, based on the fitted yield of the partially reconstructed backgrounds in the corresponding \BpmtoDpipm channel. 

In the \BtoDpi channels, a total signal yield of approximately $56\,100$ ($7750$) is found in the signal region 5249--5319\mevcc of the \DtoKsPiPi (\DtoKsKK) channel, $31\,\%$ ($32\,\%$) of which are in the long \KS category. The purity in the signal region is found to be $98.4\,\%$ ($97.7\,\%$), with the dominant background being combinatorial. In the \BtoDK channels, a total signal yield of approximately $3900$ ($530$) is found in the signal region of the \DtoKsPiPi (\DtoKsKK) channel, again finding $31\,\%$ ($32\,\%$) of the candidates in the long \KS category. The purity in the signal region is found to be $81\,\%$. The dominant background is from misidentified \BtoDpi decays, which accounts for $66\,\%$ of the background in the signal region. Equal amounts of combinatorial background and partially reconstructed decays, predominantly including a misidentified companion pion, make up the remaining background.

\section{Event selection and yield determination for \texorpdfstring{\boldmath${\TitleBztoDstmu}$ decays}{B to Dstar mu}} 
\label{sec:dmu}

A sample of \BztoDstmuX, $\Dstarpm \to \DorDbar\pipm$, $\DorDbar\to\Kshh$ decays is used to determine the quantities $F_i$, defined 
in Eq.~(\ref{eq:fi}), as the expected fractions of \Dz decays falling into the $i$th Dalitz plot bin, 
taking into account the efficiency profile of the signal decay.
The semileptonic decay of the $B$ meson and the strong-interaction decay of the \Dstarpm meson allow the flavour 
of the \Dz meson to be determined from the charges of the muon and the soft pion from the \Dstarpm decay. 
This particular decay chain, involving a flavour-tagged \Dz decay, is chosen due to its high yield, low background level, and low mistag probability. 
The selection requirements are chosen to minimise changes to the 
efficiency profile with respect to those associated with the \BpmtoDKpm channels. 

The selection is identical to that applied in Ref.~\cite{LHCb-PAPER-2014-041}, except for a tighter requirement
on the significance of the \Dz flight distance that helps to suppress backgrounds from charmless $B$ decays. To improve the resolution of the distribution of candidates across the Daltiz plot, the \B-decay chain is refitted~\cite{Hulsbergen:2005pu} with the \Dz and \KS candidates constrained to their known masses. An additional fit, in which only the \KS mass 
is constrained, is performed to improve the \Dz and \Dstarpm mass resolution in the invariant-mass fit used to determine signal yields.  


The invariant mass of the \Dz candidate, $m(\Kshh)$, and the invariant-mass difference, $\Delta m \equiv m(\Kshh\pipm)-m(\Kshh)$, are fitted 
simultaneously to determine the signal yields. 
This two-dimensional parameterisation allows the yield of selected candidates to be measured in three categories: true $\Dstarpm$ candidates (signal), candidates 
containing a true \Dz meson but random soft pion (RSP) and candidates formed from random track combinations that fall within the fit 
range (combinatorial background). Background contributions from real \Dstarpm decays paired with a random $\mu$ are determined to be negligible by selecting pairs of $\Dstarpm$ mesons and $\mupm$ with the same charge.

\begin{figure}[t]
\begin{center}
\includegraphics[width=0.45\textwidth]{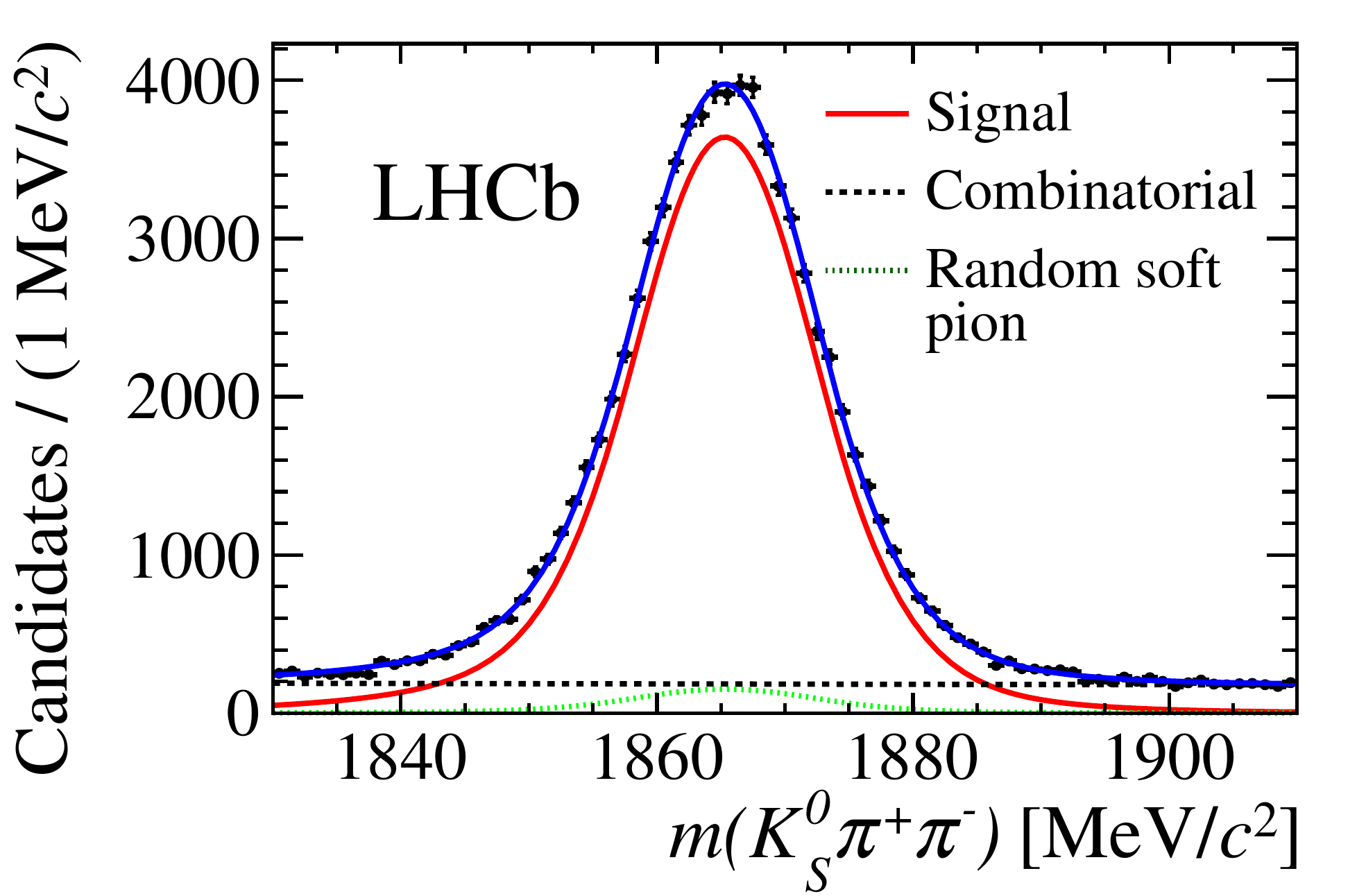} 
\includegraphics[width=0.45\textwidth]{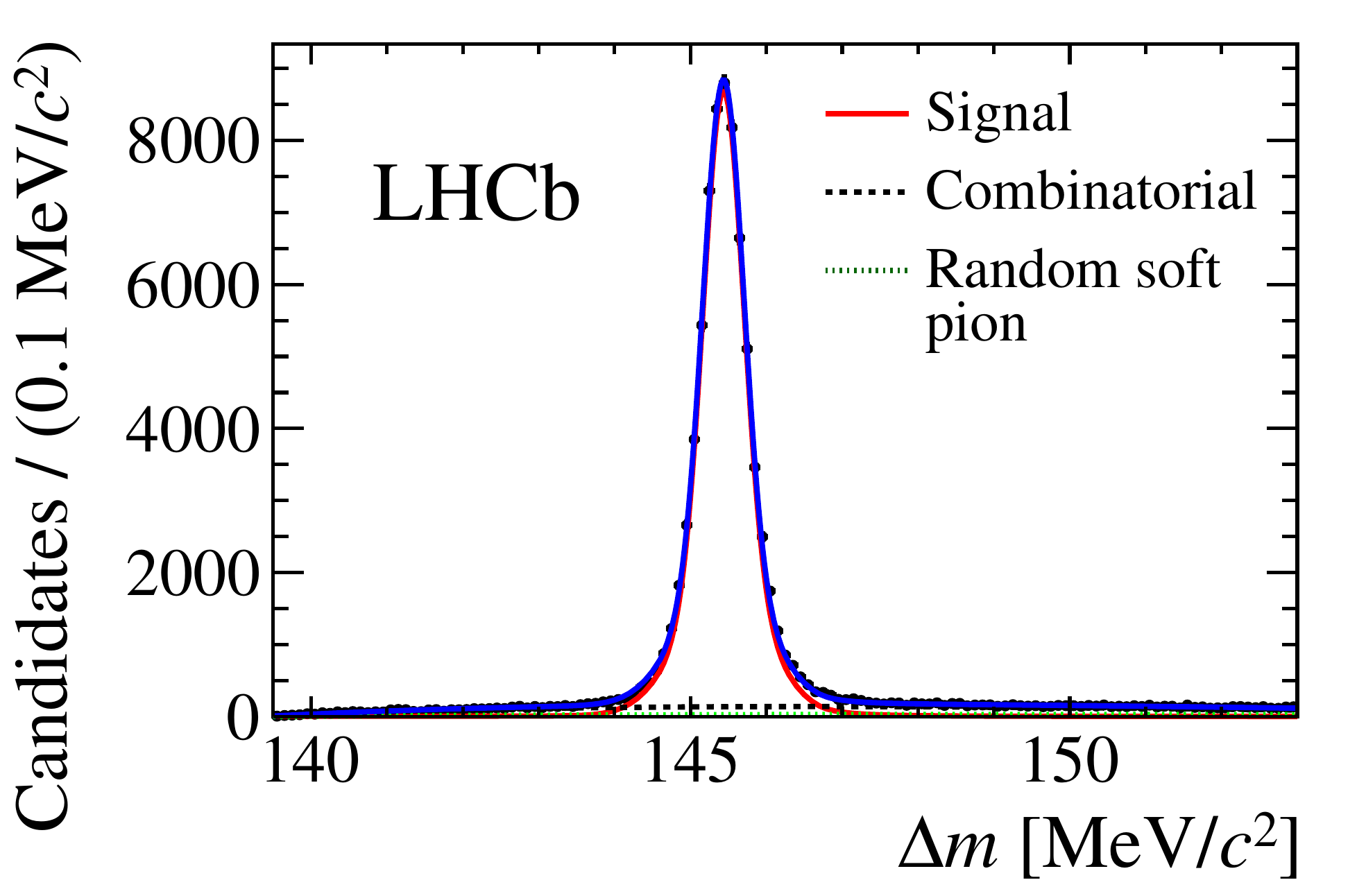} 
\caption{Result of the simultaneous fit to \BztoDstmu, $\Dstar^\pm\to\DorDbar^0(\to\KsPiPi)\pipm$ decays with downstream \KS candidates, in 2016 data. The projections of the fit result are shown for (left) $m(\KsPiPi)$ and (right) $\Delta m$. The (blue) total fit PDF is the sum of components describing (solid red) signal, (dashed black) combinatorial background and (dotted green) random soft pion background.}
\label{fig:slfit2016pipiDD}
\end{center}
\end{figure}

An example projection of $m(\KS\pip\pim)$ and $\Delta m$ is shown in Fig.~\ref{fig:slfit2016pipiDD}. The result of a two-dimensional extended, 
unbinned, maximum likelihood fit is superimposed. 
The fit is performed simultaneously for the two \Dz final states and the two \KS 
categories with some parameters allowed to be independent between categories. Candidates selected from data recorded in 2015 and 2016 are fitted separately, in order to accommodate different trigger threshold settings that result in slightly different Dalitz plot efficiency profiles. The fit region is defined by $1830 < m(\Kshh) < 1910\mevcc$ 
and $139.5 < \Delta m < 153.0\mevcc$. Within this $m(\Kshh)$ range, the $\Delta m$ resolution does not vary significantly.

The signal is parameterised in $\Delta m$ with a sum of two Crystal Ball functions, as 
for the \BpmtoDhpm signal. The mean, $\mu$, is shared between all categories, while the other parameters 
are different for long and downstream candidates. The tail parameters are fixed from simulation.
The combinatorial and RSP backgrounds are both 
parameterised with an empirical model given by
\begin{equation}
f(\Delta m ; \Delta m_0, x, p_1, p_2) =
\left [1 - \exp\left (-\frac{\Delta m - \Delta m_0}{x}\right )\right ] \left({\frac{\Delta m}{\Delta m_0}}\right )^{p_1} + p_2\left (\frac{\Delta m}{\Delta m_0} -1\right)
\end{equation}
for $\Delta m - \Delta m_0 > 0$ and $f(\Delta m)=0$ otherwise,
where $\Delta m_0$, $x$, $p_1$, and $p_2$ are free parameters. The parameter $\Delta m_0$, which describes the kinematic 
threshold for a $\Dstarpm \to \DorDbar{}^0 \pi^\pm$ decay, is shared in all data categories and for both the combinatorial and RSP shapes. 
The remaining parameters are determined separately for \DtoKsPiPi and \DtoKsKK candidates. 

In the $m(\Kshh)$ fit, all of the parameters in the signal and RSP PDFs are constrained to be 
the same as both describe a true \Dz candidate. These are also fitted with a sum of two Crystal Ball functions,
with the tail parameters fixed from simulation.
The parameters are fitted separately for the \DtoKsPiPi and \DtoKsKK shapes, due to the different phase space available in the \Dz decay. 
The combinatorial background is parameterised by an exponential function in $m(\Kshh)$. 

A total signal yield of approximately 113\,000 (15\,000) \DtoKsPiPi (\DtoKsKK) decays is obtained. This is
approximately 25 times larger than the \BtoDK yield.
In the range surrounding the signal peaks,
defined as 1840--1890 (1850--1880)\mevcc in $m(\KsPiPi)$ ($m(\KsKK)$) and 143.9--146.9\mevcc 
in $\Delta m$, the background components account for 2--5\,\% of the yield depending on the category.
 
\begin{figure}[tb]
\centering
\includegraphics[width=0.47\textwidth]{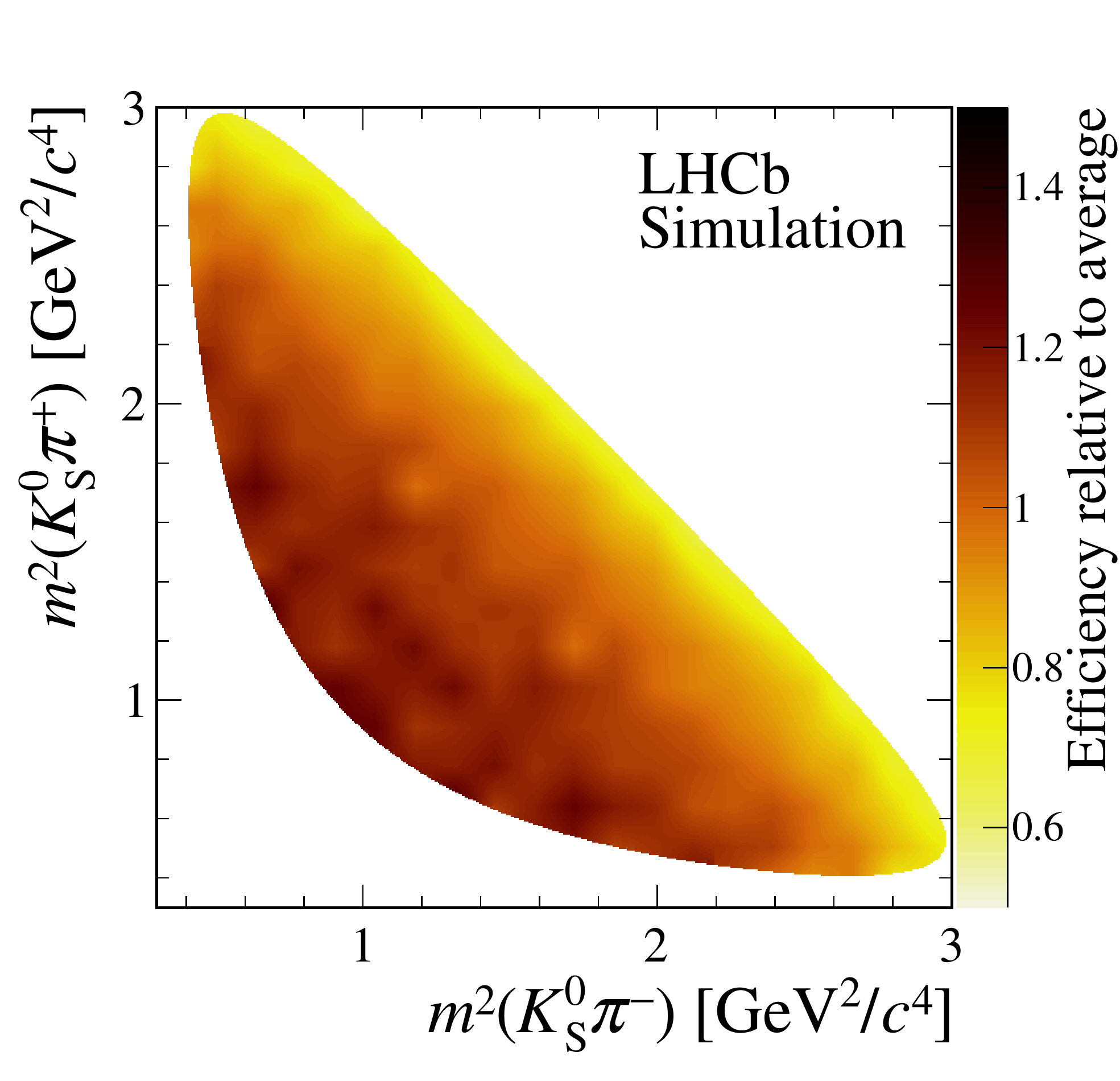}\hfill
\includegraphics[width=0.47\textwidth]{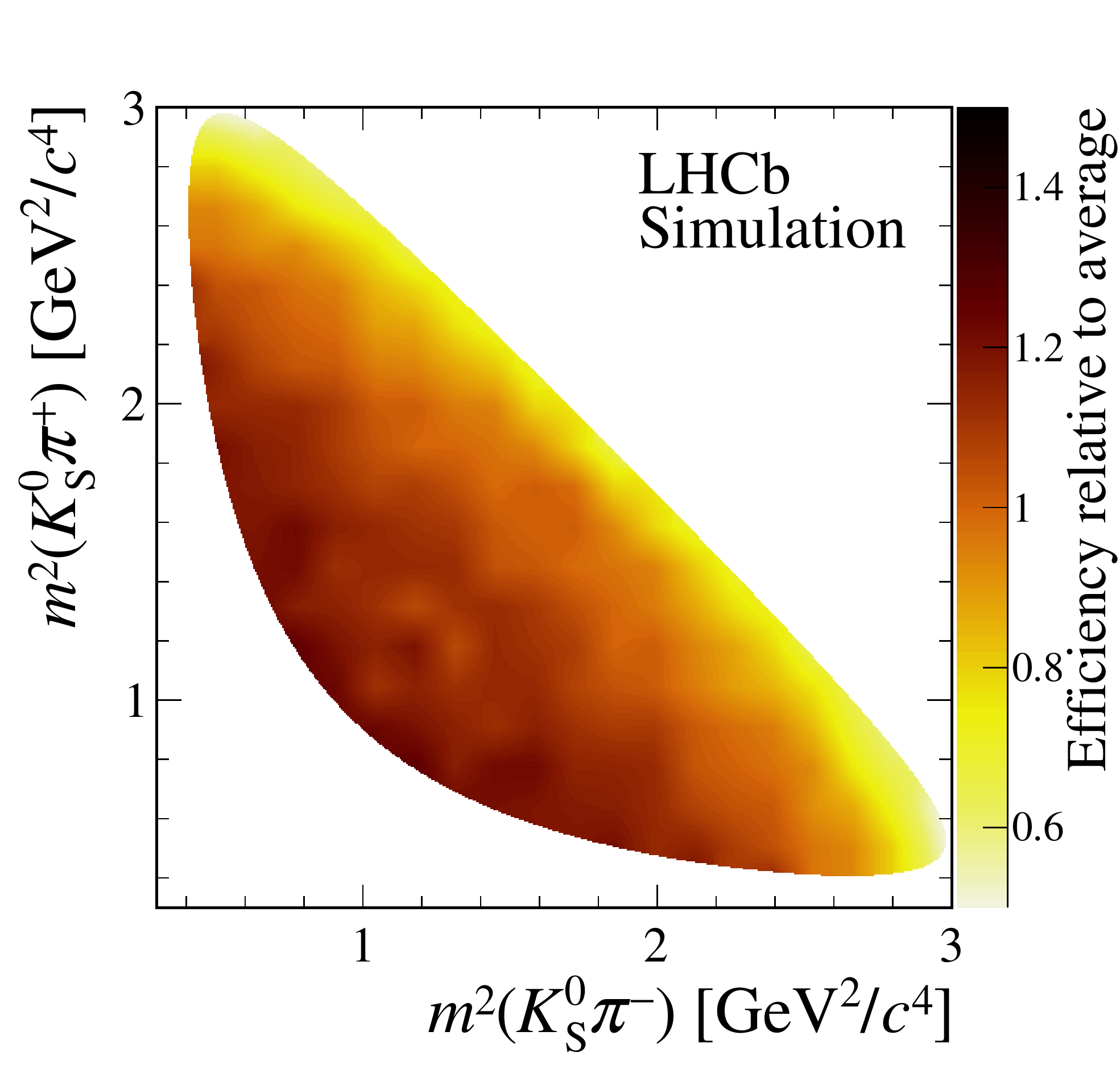} \\
\includegraphics[width=0.47\textwidth]{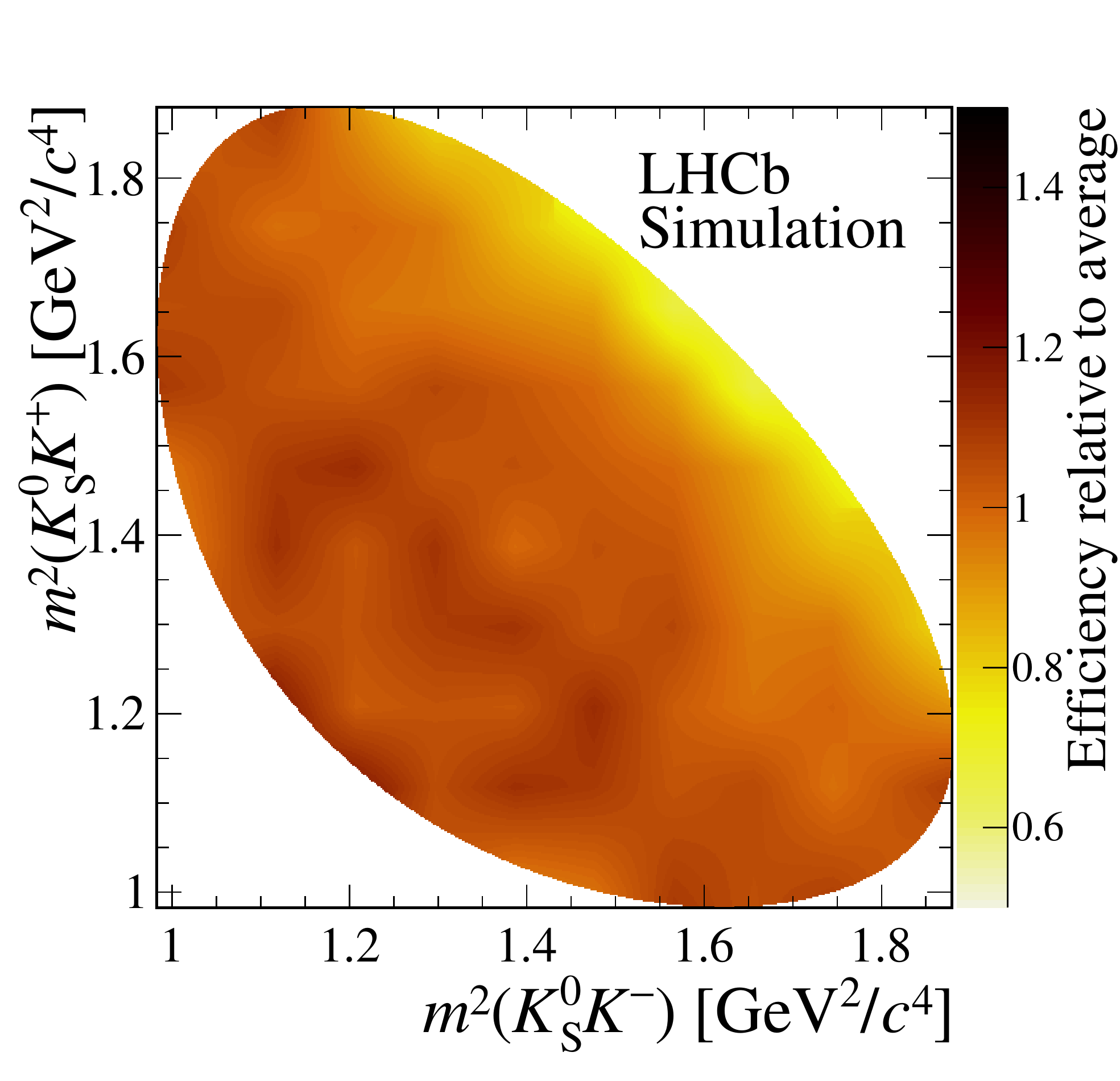}\hfill
\includegraphics[width=0.47\textwidth]{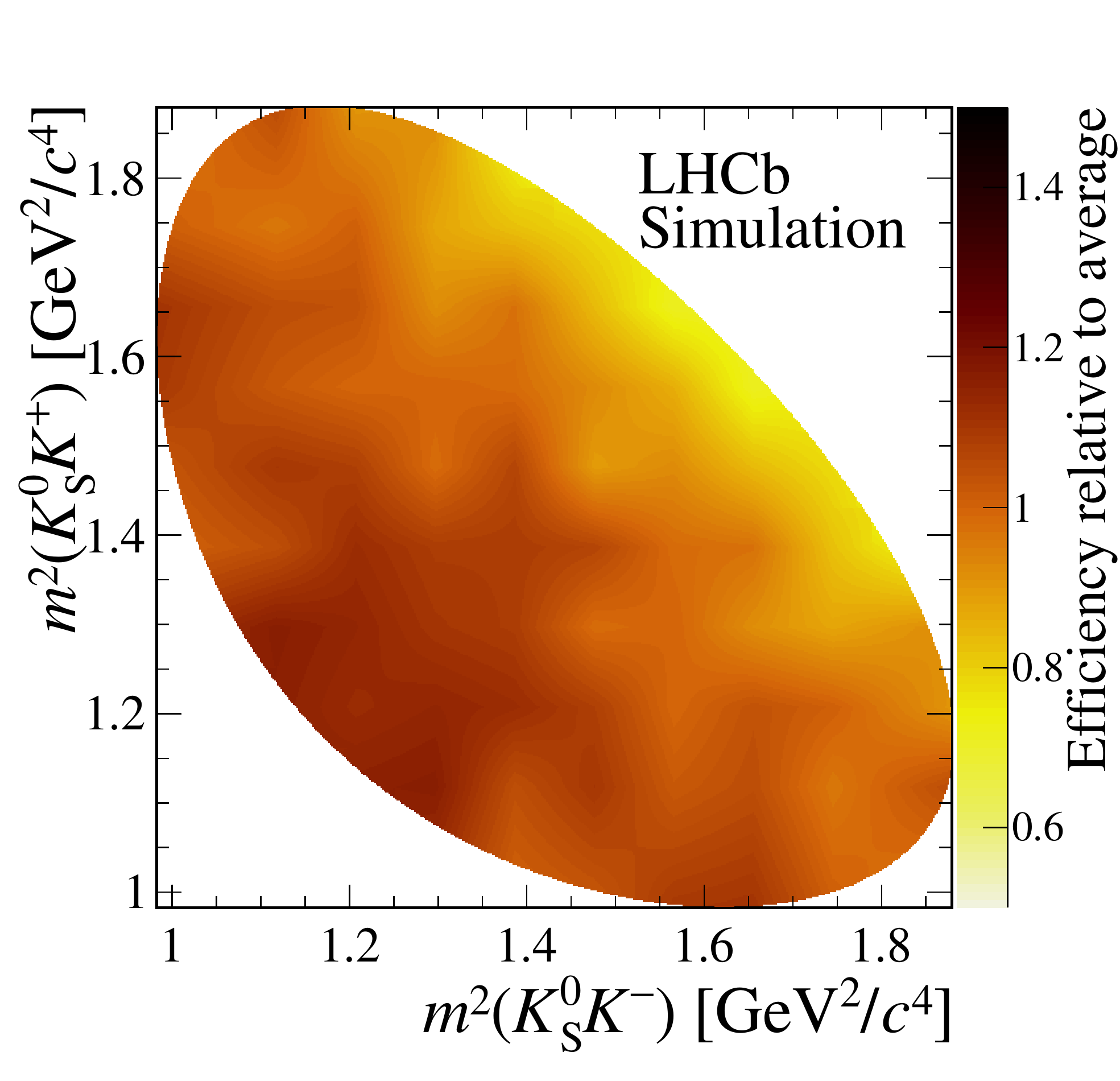}
\caption{
Example efficiency profiles of (left) \BtoDpi and (right) \BztoDstmuX decays in the simulation. The top (bottom) plots are for \DtoKsPiPi (\DtoKsKK) decays.
 These plots refer to downstream \KS candidates under 2016 data taking conditions. The normalisation is chosen so that the average over the Dalitz plot is unity. }
\label{fig:mceff}
\end{figure}

The two-dimensional fit in $m(\Kshh)$ and $\Delta m$ of the \BztoDstmuX decay is repeated in each Dalitz plot bin with all 
of the PDF parameters fixed, resulting in a raw control-mode yield, $R_i$, for each bin $i$.  
The measured $R_i$ are not equivalent to the $F_i$ fractions required to determine the \CP parameters 
due to unavoidable differences from selection criteria in the efficiency profiles of the signal and control modes.
Examples of the efficiency profiles from 
simulation of the downstream candidates in 2016 data are shown in Fig.~\ref{fig:mceff}. 
For each Dalitz plot bin $i$ a correction factor $\xi_i$ is determined to account for these efficiency differences, defined as
\begin{equation}
\xi_{i} \equiv \frac{ \int_{i} d\msqmin \, d\msqplus \, |A_{D}(\msqmin,\msqplus)|^2\,\etaDPi}{ \int_{i} d\msqmin \, d\msqplus \, |A_{D}(\msqmin,\msqplus)|^2\, \etaDst},
\end{equation}
where \etaDPi and \etaDst are the efficiency profiles of the \BtoDpi and \BztoDstmuX decays, respectively, and are determined from simulation.  
The \BtoDpi decay mode is used rather than \BtoDK as the simulation is more easily compared to the data,
due to the larger decay rate and the smaller interference between $\Bpm\to\Dz\pi^\pm$ and $\Bpm\to\Dzb\pi^\pm$ decays, compared to in the \BtoDK decay mode.
It is verified using simulation that the efficiency profiles of the \BtoDpi and \BtoDK decays are the same.
The simulated events are generated with a flat distribution across the $\D\to\KS h^+ h^-$ phase space; hence the distribution of simulated events after triggering, reconstruction and 
selection is directly proportional to the efficiency profile. 
The amplitude models used to determine the Dalitz plot intensity for the correction factor are those from Ref.~\cite{BABAR2008} and Ref.~\cite{BABAR2010} for 
the \DtoKsPiPi and \DtoKsKK decays, respectively. 
The amplitude models provide a description of the intensity distribution over the Dalitz plot and introduce no significant model dependence into the analysis.
The $F_i$ values can be determined via the relation $F_i = h'\xi_iR_i$, where $h'$ is a normalisation factor such that the sum of all $F_i$ is unity. 
The $F_i$ values are determined separately for each year of data taking and \KS category and are then combined in the fractions observed in the \BtoDpi signal region in data.  
This method of determining the $F_i$ parameters is preferable to using solely the amplitude models and \BtoDpi simulated events, since the method is 
data-driven. The amplitude models and simulation data enter the correction factor as a ratio, and thus imperfections in the simulation and the model cancel at first order. 
The average correction factor over all bins is approximately $2\,\%$ from unity and the largest correction factor is within $7\,\%$. Uncertainties on these correction factors are driven by the size of the simulation samples and are of a similar size as the corrections themselves.

\section{\boldmath Dalitz plot fit to determine the \texorpdfstring{\CP}{CP}-violating parameters \texorpdfstring{$x_{\pm}$}{x} and \texorpdfstring{$y_{\pm}$}{y}}
\label{sec:dpfit}

The Dalitz plot fit is used to measure the \CP-violating parameters $x_{\pm}$ and $y_{\pm}$, as introduced 
in Sect.~\ref{sec:principles}.  Following Eq.~\eqref{eq:populations}, these parameters 
are determined from the populations of the \Bp and \Bm Dalitz plot bins, given the external information of 
the $c_i$ and $s_i$ parameters from CLEO-c data and the values of $F_i$ from the semileptonic control decay modes. 
Although the absolute numbers of \Bp and \Bm decays integrated over the $D$ Dalitz plot have some dependence on $x_{\pm}$ and $y_{\pm}$,
the sensitivity gained compared to using just the relations in 
Eq.~\eqref{eq:populations} is negligible~\cite{Gershon} given the available sample size. 
Consequently, as stated previously, the integrated yields are not used to provide information on \xpm and \ypm and 
the analysis is insensitive to $B$ meson production and detection asymmetries.

A simultaneous fit is performed on the \BpmtoDhpm data, split into the two $B$ charges, 
the two \KS categories, the $\Bpm \to D \Kpm$ and $\Bpm \to D \pipm$ candidates, and the two \DtoKshh  final states. 
The invariant mass of each \Bpm candidate is calculated using the results of a kinematic fit in which the \D and \KS masses are constrained to their known values.
Each category is then divided into the Dalitz plot bins shown in Fig.~\ref{fig:bins}, where there are 16 bins 
for \DtoKsPiPi and 4 bins for \DtoKsKK. 
The \BtoDK and \BtoDpi samples are fitted simultaneously because the yield of \BtoDpi signal in each Dalitz 
plot bin is used to determine the yield of misidentified candidates in the corresponding \BtoDK Dalitz plot bin.   
The PDF parameters for both the signal and background invariant-mass distributions are fixed to the values 
determined in the invariant-mass fit described in Sect.~\ref{sec:selection}. The $\Bpm$ mass range is reduced 
to 5150--5800\gevcc to avoid the need of a detailed description of the shape of the partially reconstructed background.  
The yields of signal candidates for each bin in the $\Bpm\to D \pipm$ sample are free parameters. 
In each of the $\Bpm\to D \Kpm$ channels, the \emph{total} yield integrated over the Dalitz plot is a free parameter. The fractional yields in each bin are defined using the expressions for the Dalitz plot distribution in terms of $\xpm,\, \ypm,\, F_i,\, c_i$, and $s_i$ in Eq.~\eqref{eq:populations}, where the $\xpm$ and $\ypm$ parameters are free and the values of $F_i$ are Gaussian-constrained within their 
uncertainties. The values of $c_i$ and $s_i$ are fixed to their central values, which is taken into account as a source of systematic uncertainty. 
The yields of the component due to \BtoDpi decays, where the companion has been misidentified as a kaon, are fixed in each \BtoDK bin, relative to the yield in the corresponding \BtoDpi bin, using the known PID efficiencies. A component for misidentified \BtoDK decays in the \BtoDpi channels is not included, as it is found to contribute less than $0.5\%$ of the yield in the signal region in the global fit described in Sect.~\ref{sec:selection}. 
The total yield of the partially reconstructed \Bpm and \Bz backgrounds is fitted in each bin, using the same shape in all bins, with the fractions of each component taken from the global fit. The total yield of the $\Bs\to\Dzb\pip\Km$ ($\Bsb\to\Dz\pim\Kp$) background is fixed in each channel, using the results of the global fit. The yield in each bin is then fixed from the $F_i$ parameters, using the known Dalitz distribution of $\Dzb(\Dz)\to\Ks h^+h^-$ decays. The separate treatment of the partially reconstructed background from \Bs decays is necessary due to the significantly different Dalitz distribution, arising because only a \Dzb meson is produced along with a \Km meson, while for the remaining modes, the \D meson is either a \Dz meson or an admixture where the \Dzb component is $r_B$-suppressed. The yield of the combinatorial background in each bin is a free parameter. In bins in which an auxiliary fit determines the yield of the partially reconstructed or combinatorial background to be negligible, the corresponding yields are set to zero to facilitate the calculation of the covariance matrix~\cite{MINUIT,James:1975dr}.  

A large ensemble of pseudoexperiments is performed to validate the fit procedure.  In each pseudoexperiment 
the numbers and distributions of signal and background candidates are generated according to the expected distribution  
in data, and the full fit procedure is then executed.  The input values for $x_{\pm}$ and $y_{\pm}$ correspond to $\gamma = 70^\circ$, $r_{B}=0.1$,  
and $\delta_B = 130^\circ$. The uncertainties determined by the fit to data are 
consistent with the size of the uncertainties determined by the pseudoexperiments. Small biases are observed
in the central values and are due to the low event yields in some of the bins. These biases are observed 
to decrease in simulated experiments of larger size. The central values are corrected for the biases
and a systematic uncertainty is assigned, as described in Sect.~\ref{sec:syst}.

The \CP parameters obtained from the fit are
\begin{align*}
    \xm & = (\phantom{-}9.0 \pm 1.7 ) \times 10^{-2}\,, \\
    \ym & = (\phantom{-}2.1 \pm 2.2 ) \times 10^{-2}\,, \\
    \xp & = (         - 7.7 \pm 1.9 ) \times 10^{-2}\,, \\
    \yp & = (         - 1.0 \pm 1.9 ) \times 10^{-2}\,,
\end{align*}
where the uncertainties are statistical only. The correlation matrix is shown in Table~\ref{tab:stat_correlations}. The total \BtoDK yields in the signal region, where the invariant mass of the \B candidate is in the interval 5249--5319\mevcc, are shown in Table~\ref{tab:yields_fit}. 
\begin{table}[tb]
    \centering
    \renewcommand\arraystretch{1.2}
    \caption{
    Statistical correlation matrix for the fit to data.
    \label{tab:stat_correlations}
    }
    \begin{minipage}{0.45\linewidth}
        \begin{tabularx}{\textwidth}{l|YYYY}
        \hline
        & $\phantom{-}\xm$ & $\phantom{-}\ym$ & $\phantom{-}\xp$ & $\phantom{-}\yp$ \\
        \hline
        \xm & $\phantom{-}1$ & $-0.21$ & $\phantom{-}0.05$ & $\phantom{-}0.00$ \\
        \ym & & $\phantom{-}1$ & $-0.01$ & $\phantom{-}0.02$ \\
        \xp & & & $\phantom{-}1$ & $\phantom{-}0.02$ \\
        \yp & & & & $\phantom{-}1$\\
        \hline
        \end{tabularx}

        
    \end{minipage}
    
\end{table}

\begin{table}
    \centering
    \caption{Fit results for the total $\Bpm \to \D\Kpm$ yields in the signal region, where the invariant mass of the \B candidate is in the interval 5249--5319\mevcc, integrated over the Dalitz plots.
    \label{tab:yields_fit}
    }
    \begin{tabular}{lccccc}
        \hline
         & \multicolumn{2}{c}{$\Bm \to \D\Km$}  &  & \multicolumn{2}{c}{$\Bp \to \D\Kp$}       \\
        \cline{2-6}
                            & Long              & Downstream        && Long             & Downstream        \\
        \hline
        $\D\to\KS\pip\pim$  & $602 \pm 26$   & $1\,315\pm39$      && $606\pm26$        & $1\,334\pm39        $\\
        $\D\to\KS\Kp\Km$    & $\phantom{6}92\pm10 $ & $\phantom{1\,}189\pm15$       && $\phantom{6}82\pm10 $& $ \phantom{1\,}193 \pm 15 $ \\
        \hline 
    \end{tabular}
\end{table}

\begin{figure}[tb]
\begin{center}
\includegraphics[width=0.55\textwidth]{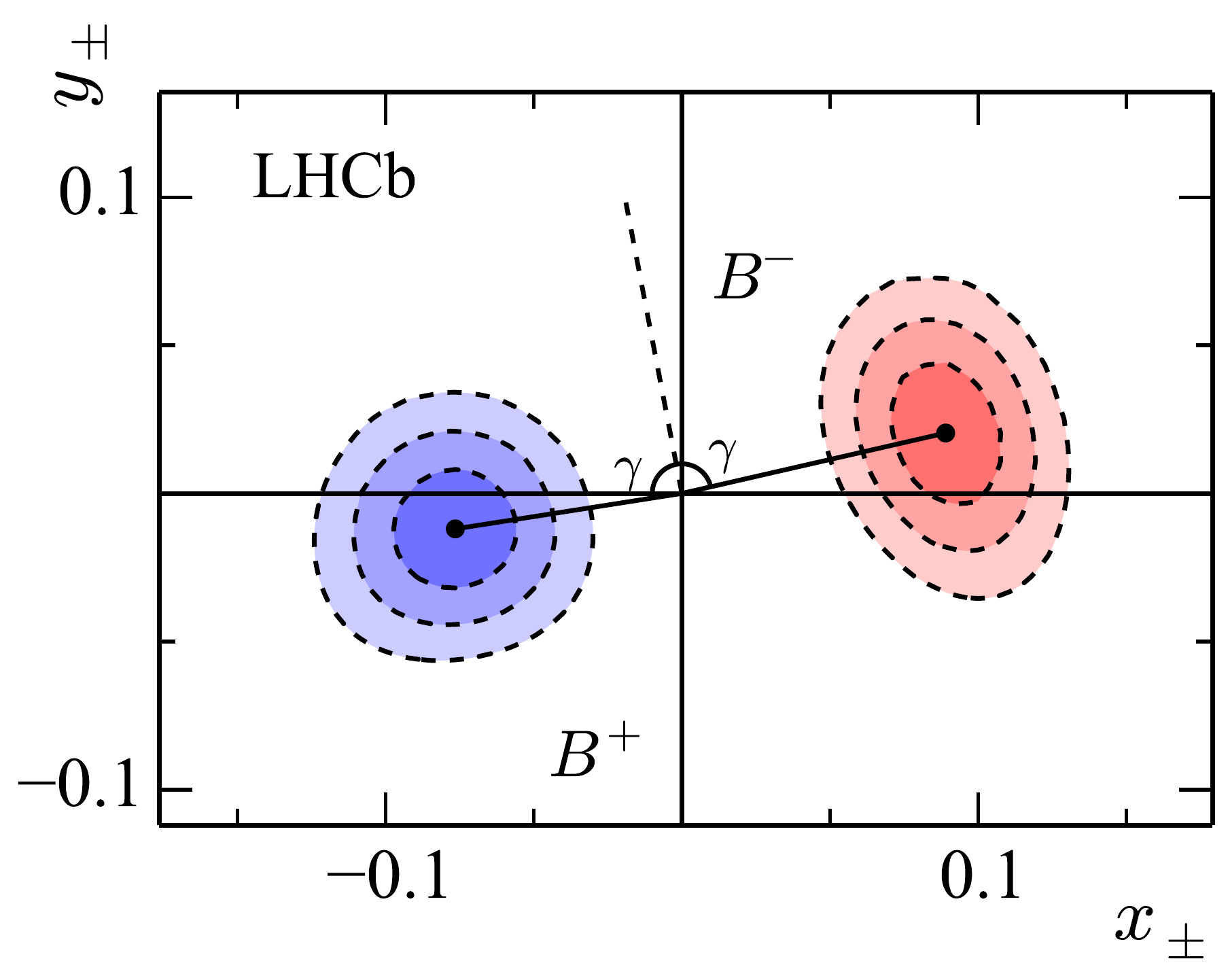}
\caption{\small Confidence levels at 68.2\%, 95.5\% and 99.7\% probability for 
$(x_+,y_+)$ and $(x_{-}, y_{-})$ as measured in $B^\pm \to D K^\pm$ decays (statistical uncertainties only). The parameters $(x_+,y_+)$ relate to \Bp decays and $(x_{-}, y_{-})$ refer to \Bm decays. The black dots show the central values obtained in the fit.}
\label{fig:sunnysideup}
\end{center}
\end{figure}

The measured values of $(x_{\pm}, y_{\pm})$ from the fit to data are displayed in Fig.~\ref{fig:sunnysideup}, along with their likelihood contours, 
corresponding to statistical uncertainties only. The systematic uncertainties are discussed in the next section. 
The two vectors defined by the coordinates $(x_-,y_-)$ and $(x_+,y_+)$ are not consistent with zero magnitude and they have a non-zero opening angle. Therefore the data sample exhibits the expected features of \CP violation. The opening angle is equal to $2\gamma$, as illustrated in Fig.~\ref{fig:sunnysideup}. 

\begin{figure}[t]
    \centering
    \includegraphics[width=0.48\textwidth]{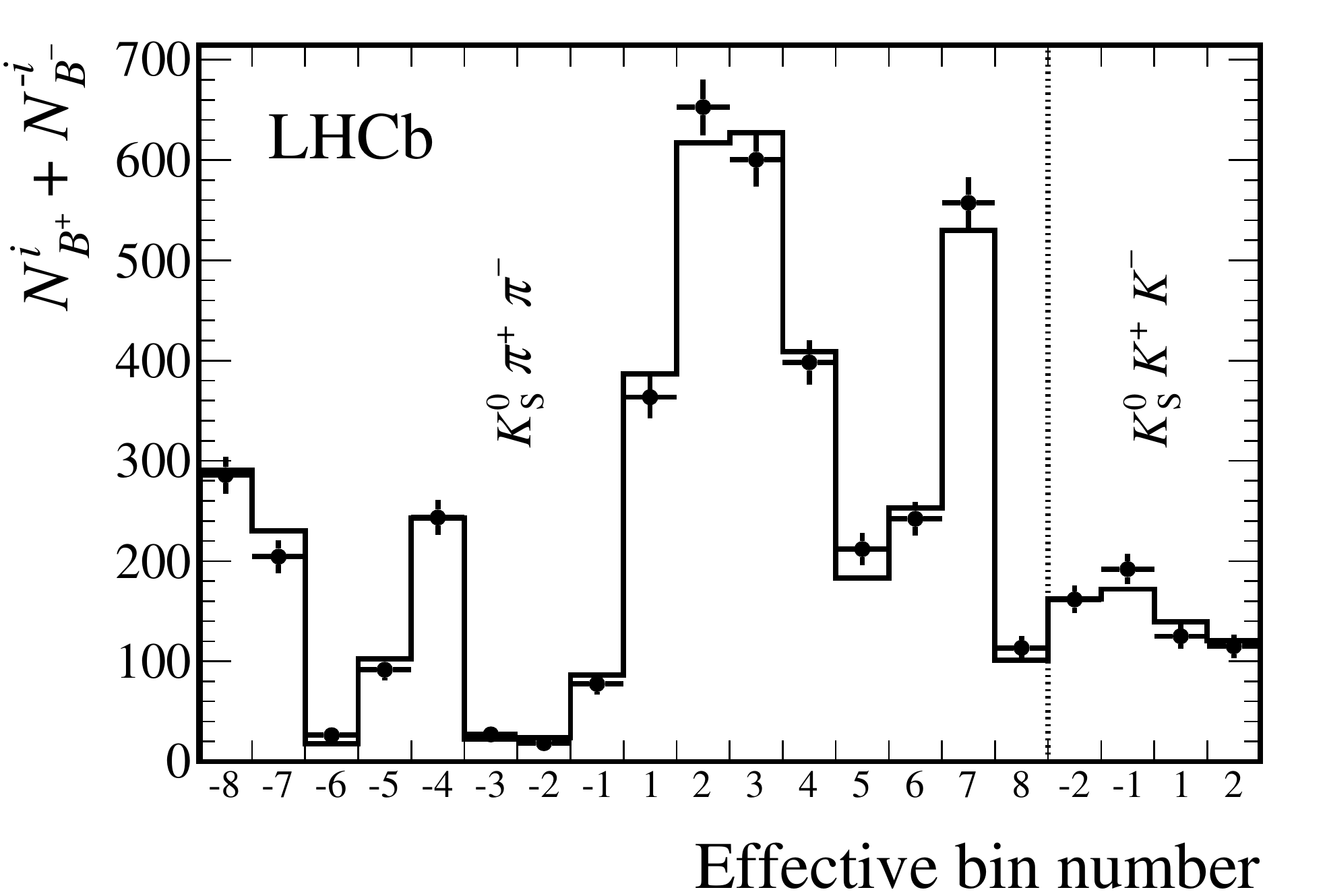}
    \includegraphics[width=0.48\textwidth]{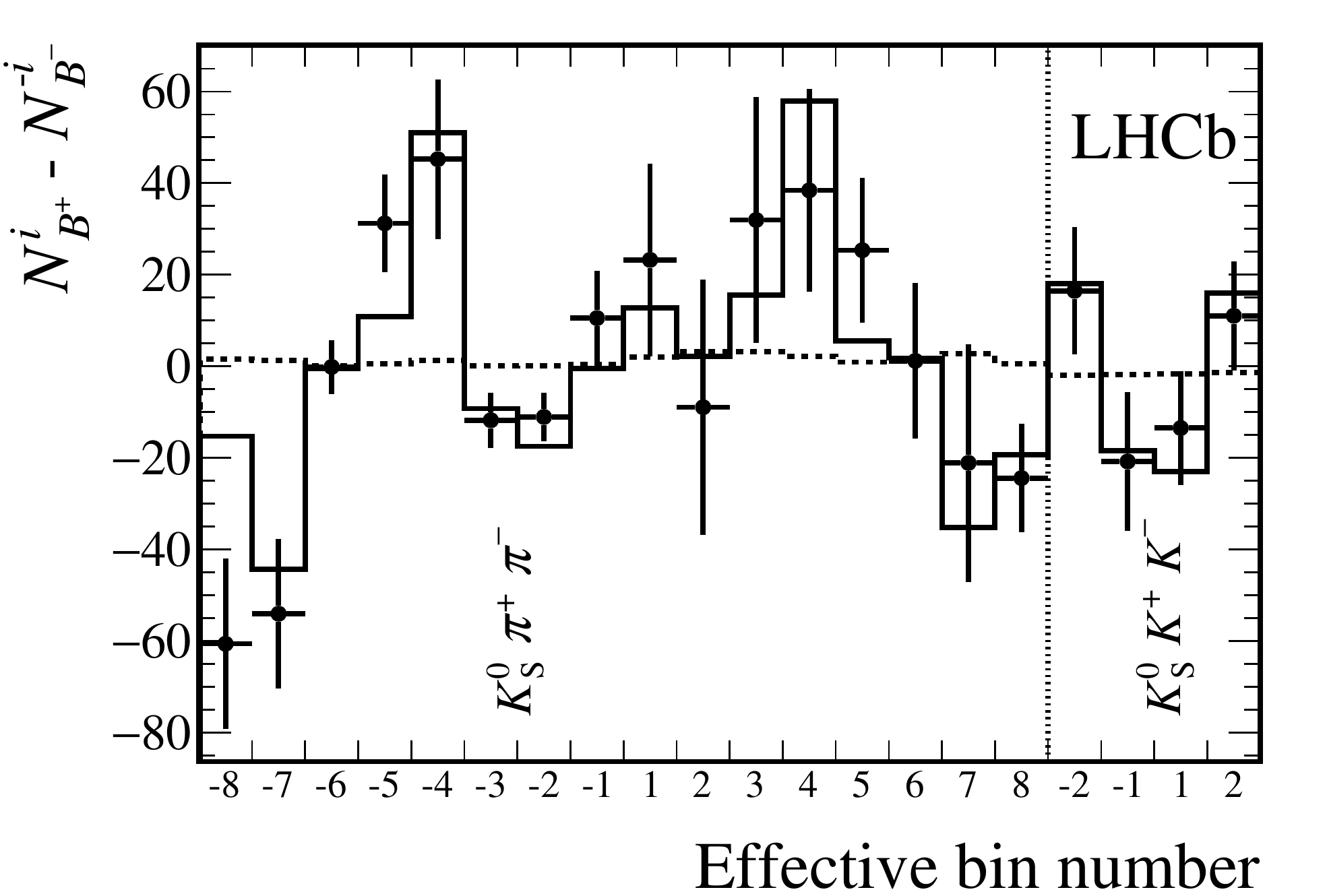}
    \caption{(Left) Comparison of total signal yields from the direct fit (points) to those calculated from the central values of $x_\pm$ and $y_\pm$ (solid line). The yields are given for the effective bin: $+i$ for \Bp and $-i$ for \Bm, and summed over \B charge and \Ks decay category. (Right) Comparison of the difference between the \Bp and \Bm yield obtained in the direct fit for each effective bin (points), the prediction from the central values of $x_\pm$ and $y_\pm$ (solid line), and the prediction assuming no \CP violation (dotted line).}
    \label{fig:yield_and_diff_comparison}
\end{figure}

In order to assess the goodness of fit, and to demonstrate that the equations in $(x_\pm,y_\pm)$ provide a good description of data, an alternative fit is performed where the \BtoDK yields are measured independently in each bin. In Fig.~\ref{fig:yield_and_diff_comparison} (left) the obtained yields are compared with the yields predicted from the values of $(x_\pm,y_\pm)$ obtained in the default fit. The yields from the direct fit agree with the prediction with a $p$-value of 0.33. In Fig.~\ref{fig:yield_and_diff_comparison} (right) the difference $N_\Bp^i - N_\Bm^{-i}$ in each bin is calculated using the results of the direct fit of the \BtoDK yields. This distribution is compared to that predicted by the central $(x_\pm,y_\pm)$ values. The measured yield differences are compatible with the prediction with a $p$-value of 0.58. In addition, data are fitted with the assumption of no \CP violation by enforcing $\xp=\xm\equiv x_0$ and $\yp=\ym\equiv y_0$. The obtained $x_0$ and $y_0$ values are used to determine the predicted values of $N_\Bp^i - N_\Bm^{-i}$, which are also shown in Fig.~\ref{fig:yield_and_diff_comparison} (right). This prediction is not zero because the \B meson production and various detection effects can induce a global asymmetry in the measured yields. The comparison of the data to this hypothesis yields a $p$-value of $1\times 10^{-6}$, which strongly disfavours the \CP-conserving hypothesis.

\section{Systematic uncertainties}
\label{sec:syst}

Systematic uncertainties on the measurements of the \xpm and \ypm parameters are evaluated and are presented in Table~\ref{tab:syst_all}. 
The source of each systematic uncertainty is described below. The systematic uncertainties are generally
determined from an ensemble of pseudoexperiments where the simulated data are generated in an alternative configuration and 
fitted with the default method. The mean shifts in the fitted values of \xpm and \ypm in 
comparison to their input values are taken as the systematic uncertainty.

\begin{table}[tb]
\begin{center}
\caption{\small 
Summary of uncertainties for the parameters \xpm and \ypm. The various sources 
of systematic uncertainties are described in the main text.
All entries are given in multiples of $10^{-2}$.}\label{tab:syst_all} 
\vspace*{0.1cm}
\begin{tabular}{l c c c c }
\hline
Source & $\quad x_-\quad$ & $\quad y_-\quad$ & $\quad x_+\quad$ & $\quad y_+\quad$ \\
\hline 
Statistical                                    & 1.7 & 2.2    & 1.9 & 1.9   \\
\hline
Strong phase measurements                      & 0.4 & 1.1    & 0.4 & 0.9   \\
\hline
Efficiency corrections                         & 0.6 & 0.2    & 0.6 & 0.1   \\
Mass fit PDFs                                  & 0.2 & 0.3    & 0.2 & 0.3    \\
Different mis-ID shape over Dalitz plot         & 0.2 & 0.1    & 0.1 & 0.1   \\
Different low mass shape over Dalitz plot      & 0.1 & 0.2    & 0.1 & 0.1   \\
Uncertainty on $\Bs\to\Dzb\pip\Km$ yield     & 0.1 & 0.1    & 0.1 & 0.1   \\
Bias correction                                & 0.1 & 0.1    & 0.1 & 0.1   \\
Bin migration                                      & 0.1 & 0.1    & 0.1 & 0.1   \\
$K^0$ \CP violation and material interaction             & 0.1 & 0.2    & 0.1 & 0.1   \\
\hline
Total experimental systematic uncertainty                      & 0.7 & 0.5    & 0.7 & 0.4    \\
\hline
\end{tabular}
\end{center}
\end{table}

The limited precision on $(c_i, s_i)$ coming from the CLEO measurement induces uncertainties on $x_\pm$ and $y_\pm$~\cite{CLEOCISI}.
These uncertainties are evaluated by 
fitting the data multiple times, each with different $(c_i, s_i)$ values sampled according to their experimental uncertainties and correlations. 
The resulting widths in the distributions of \xpm and \ypm values are assigned as the systematic uncertainties. Values of $(0.4\text{--}1.1) \times 10^{-2}$ 
are found for the fit to the full sample. The uncertainties are similar to, but different from, those reported 
in Ref.~\cite{LHCb-PAPER-2014-041}. This is as expected since 
it is found from simulation studies that the \linebreak($c_i$, $s_i$)-related uncertainty depends on the particular sample under study. It is found that the uncertainties do become constant when simulated samples with very high signal yields are studied. The uncertainties arising from the CLEO measurements are kept separate from the other experimental uncertainties. 

\begin{figure}[tb]
\centering
\includegraphics[width=0.47\textwidth]{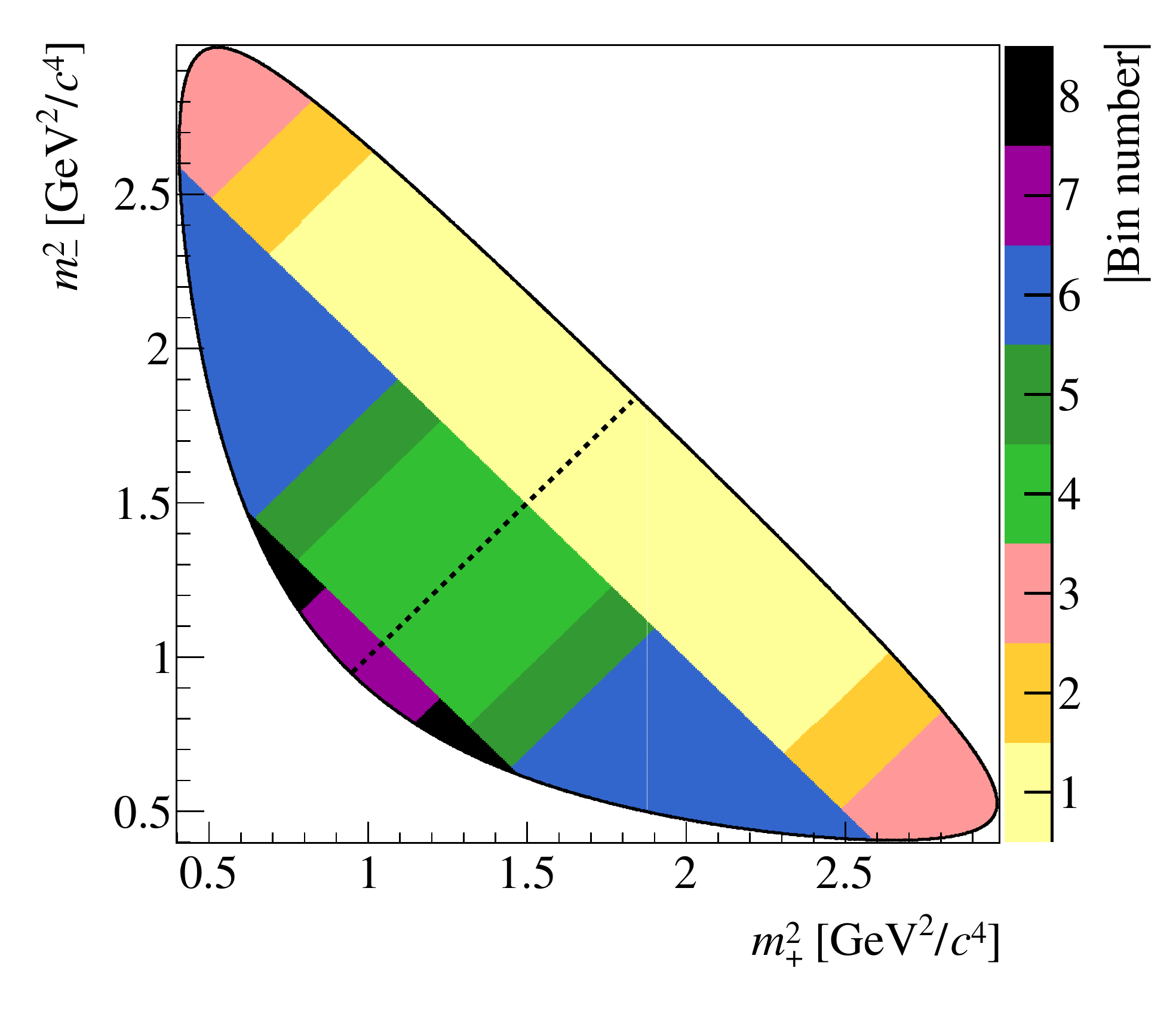}
\includegraphics[width=0.47\textwidth]{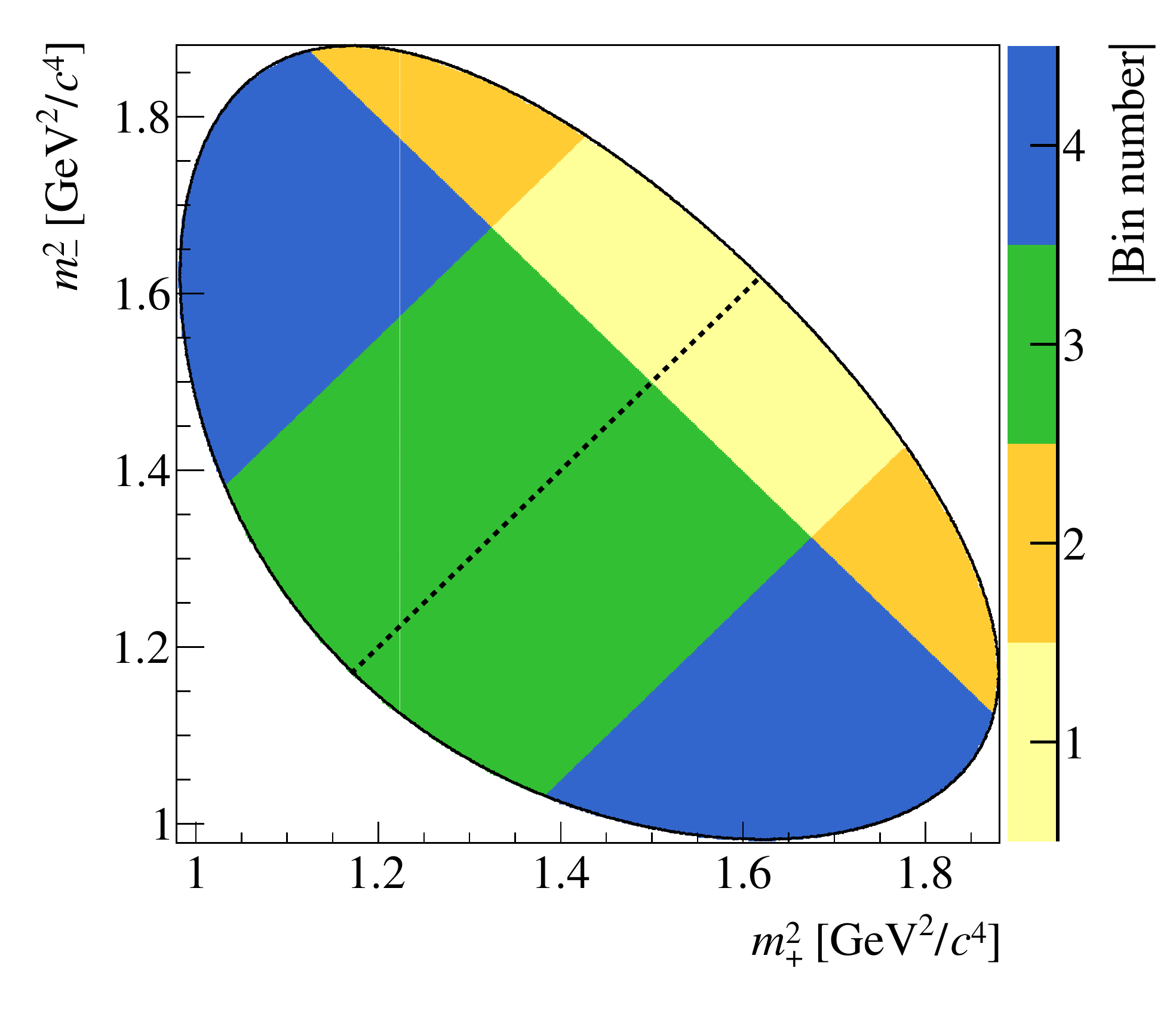}
\caption{Rectangular binning schemes for (left) \DtoKsPiPi decays and (right) \DtoKsKK decays. The diagonal line separates the positive and negative bins, where the positive bins are in the region in which $m^2_->m^2_+$ is satisfied. \label{fig:rect}}

\end{figure}

A systematic uncertainty arises from imperfect modelling in the simulation used to derive the efficiency 
correction for the determination of the $F_i$ parameters. 
As the simulation enters the correction in a ratio, it is expected that imperfections cancel to first order.
To determine the residual systematic uncertainty associated with this correction, 
an additional set of correction factors is calculated and used to evaluate an alternative 
set of $F_i$ parameters. To determine this additional factor, a new rectangular binning scheme is used, which is shown in Fig.~\ref{fig:rect}. 
The bin-to-bin efficiency variation in this rectangular scheme is significantly larger than for the default 
partitioning and is more sensitive to imperfections in the simulated data efficiency profile. 
The yields of the \BtoDpi and \BztoDstmuX decays in each bin of the rectangular scheme are compared to the predictions from the amplitude model 
and the simulated data efficiency profile. The usage of the rectangular binning also  helps to dilute the small level of \CP violation in \BtoDpi
such that differences from this comparison will come primarily from efficiency effects.
The alternative correction factors $\xi_{i}^{ \textrm{alt}}$ are calculated as
\begin{equation}
\label{eq:altcor}
\xi_{i}^{ \textrm{alt}} =\frac{ \int_{i} d\msqmin d\msqplus \etaDPi\, |A_{D}(\msqmin,\msqplus)|^2 \, C_{D\pi}(\msqmin,\msqplus)}{ \int_{i} d\msqmin d\msqplus \etaDst\, |A_{D}(\msqmin,\msqplus)|^2 \, C_{\Dstar\!\mu}(\msqmin,\msqplus)}\,,
\end{equation}
where the $C(m^2_-,m^2_+)$ terms are the ratios between the predicted and observed data yields in the rectangular bins. Many pseudoexperiments 
are performed, in which the data are generated according to the alternative $F_i$ parameters and then fitted with the default $F_i$ parameters. 
The overall shift in the fitted values of the \CP parameters in comparison to their input values is taken as 
the systematic uncertainty, yielding $0.6\times 10^{-2}$ for \xpm and $0.1 (0.2)\times 10^{-2}$ for \yp (\ym).

Various effects are considered to assign an uncertainty for the imperfections in the description of the invariant-mass spectrum. For the PDF used to fit the signal, the parameters of the PDF used in the binned fit are varied according to the uncertainties obtained in the global fit. An alternative shape is also tested. 
The global fit is repeated with the mean and width of the shape used to describe the background due to misidentified companions allowed to vary freely. The results are used to generate data sets with an alternative PDF, and fit them using the default setup.
The description of the partially reconstructed background is changed to a shape obtained from a fit of the PDF to simulated decays. 
The slope of the exponential used to fit the combinatorial background is also fluctuated according to the uncertainty
obtained in the global fit.
The contributions from each change are summed in quadrature and are $0.2 \times 10^{-2}$ for each of the \xpm parameters and $0.3 \times 10^{-2}$ for each of the \ypm parameters.

Two systematic uncertainties associated with the misidentified \BpmtoDpipm background in the \BpmtoDKpm sample are considered. 
First, the uncertainties on the particle misidentification probabilities are found to have a negligible effect on the measured values of \xpm and \ypm. 
Second, it is possible that the invariant-mass distribution of the misidentified background (the mis-ID shape) is not uniform over the Dalitz plot, as assumed in 
the fit. This can occur through kinematic correlations between  the reconstruction efficiency across the Dalitz plot of the $D$ decay 
and the momentum of the companion pion from the \Bpm decay. Alternative mass shapes are constructed by repeating the procedure used to obtain the default shape for each Dalitz bin individually. The alternative
shapes are used when generating data sets for pseudoexperiments, and the fits then performed assuming 
a single shape, as in the fit to data. The resulting uncertainty is at most $0.2 \times 10^{-2}$ for all \CP parameters. 

In the fit to the data, the relative contributions of the partially reconstructed \Bpm and \Bz backgrounds
are kept the same in each Dalitz bin. This is a simplification as some partially reconstructed backgrounds will
be distributed as $\Dz(\Dzb)\to\KS h^+h^-$ for reconstructed \Bm (\Bp) candidates, while partially reconstructed $\B^\pm \to D^{(*)}K^{(*)\pm}$ decays will be distributed as a $\Dz-\Dzb$ admixture depending on the relevant \CP violation parameters. Pseudoexperiments are generated, where the $D$-decay Dalitz plot distribution for $\B^\pm \to D^*K^+$ is based on the \CP parameters reported in Ref.~\cite{LHCb-CONF-2017-004} and those for $\B^\pm \to D K^{*+}$ are taken from Ref.~\cite{LHCb-PAPER-2017-030}. The generated samples are fitted with the standard method.
The resulting uncertainty is at most $0.2 \times 10^{-2}$ for all \CP parameters. 

The total yield of the $\Bs\to\Dzb\pip\Km$ background in the \BtoDK channels  is fixed relative to the corresponding \BtoDpi yield. The systematic uncertainty due to the uncertainty on the relative rate is estimated via pseudoexperiments, where data sets are generated with the rate varied by $\pm 1\sigma$ and fitted using the default value. The maximal mean bias for each parameter is taken as the uncertainty. The resulting uncertainty is $0.1 \times 10^{-2}$ for all \CP parameters. 

An uncertainty is assigned to each \CP parameter to accompany the correction that is applied  for the small  bias observed in the fit procedure.  
These uncertainties are determined by performing sets of pseudoexperiments, each generated with different values of \xpm and \ypm 
throughout a range around the values predicted by the world averages.
The spread in observed bias is combined in quadrature with the uncertainty in the precision of the pseudoexperiments. 
This is taken as the systematic uncertainty and is $0.1 \times 10^{-2}$ for all \CP parameters.

The systematic uncertainty from the effect of candidates being assigned the wrong Dalitz bin number is considered. 
The resolution in $m^2_+$ and $m^2_-$ is approximately 0.006\gevgevcccc for 
candidates with long \KS decays and 0.007\gevgevcccc for candidates with downstream \KS decays. While this is small compared to the typical 
width of a bin, net migration can occur in regions where the presence of resonances cause the density to change rapidly.
To first order this effect is accounted for by use of the control channel. However, differences in the distributions of the Dalitz plots
due to efficiency differences or the nonzero value of $r_B$ in the signal decay may cause residual effects.
The uncertainty from this is determined via pseudoexperiments, in which different 
input $F_i$ values are used to reflect the residual migration. The size of any possible bias is found to be
$0.1 \times 10^{-2}$ for all \CP parameters.

There is a systematic uncertainty related to \CP violation in the neutral kaon system due to the fact that  the \Ks state is not an exact \CP eigenstate and, separately, due to different nuclear interaction cross-sections of the \Kz and \Kzb mesons. The measurement is insensitive to global asymmetries, but is affected by the different Dalitz distributions of $\D\to\KS h^- h^+$ and $\D\to\KL h^- h^+$ decays, as well as any correlations between Dalitz coordinates and the net material interaction.
The potential bias on \xpm and \ypm is assessed using a series of pseudoexperiments, where data are generated taking the effects into account and fitted using the default fit. The $\D\to\KL h^- h^+$ Dalitz distribution is estimated by transforming an amplitude model of $\D\to\KS h^- h^+$ \cite{BABAR2005}, following arguments and assumptions laid out in Ref.~\cite{CLEOCISI}. The effect of material interaction is treated using the formalism described in Ref.~\cite{LHCb-PAPER-2014-013}. The size of the potential bias is found to be $\leq 0.2 \times 10^{-2}$ for all \CP parameters, corresponding to a bias on $\gamma$ of approximately $0.8^\circ$, which is within expected limits \cite{Yuval}.

The nonuniform efficiency profile over the Dalitz plot means that the values of $(c_i, s_i)$ appropriate for this analysis can 
differ from those measured by the CLEO collaboration, which correspond to the constant-efficiency case. 
Amplitude models are used to calculate the values of $c_i$ and $s_i$ 
both with and without the efficiency profiles determined from simulation. The models are taken from Ref.~\cite{BABAR2008} for \DtoKsPiPi decays and from Ref.~\cite{BABAR2010} for \DtoKsKK decays. The difference is taken as an estimate of the size
of this effect. Pseudoexperiments are generated in which the values have been shifted by this difference, and then fitted with the default
 $(c_i, s_i)$ values. The resulting bias on \xpm and \ypm is found to be negligible.

The effect that a detection asymmetry between hadrons of opposite charge can have on the symmetry of the efficiency across the Dalitz plot is found 
to be negligible. Changes in the mass model used to describe the semileptonic control sample are also found to have a negligible effect on the $F_i$ values.
 
Finally, several checks are conducted to assess the stability of the results.   These include repeating the fits separately for both \KS categories, for each data-taking year, and by splitting the candidates depending on whether the hardware trigger decision was due to particles in the signal-candidate decay chain or other particles produced in the $pp$ collision. 
No anomalies are found and no additional systematic uncertainties are assigned. 

In total the systematic uncertainties are less than half of the corresponding statistical uncertainties. The correlation matrix obtained 
for the combined effect of the sources of experimental and strong-phase related systematic uncertainties is given in Table~\ref{tab:syst_correlations}.

\begin{table}[t]
    \centering
    \renewcommand\arraystretch{1.2}

    \caption{
    Correlation matrix of the experimental and strong-phase related systematic uncertainties.
    \label{tab:syst_correlations}
    }

        \begin{tabularx}{0.6\textwidth}{l|YYYY}
        \hline
        & $\phantom{-}\xm$ & $\phantom{-}\ym$ & $\phantom{-}\xp$ & $\phantom{-}\yp$ \\
        \hline
        \xm & $\phantom{-}1$ & $-0.25$ & $\phantom{-}0.43$       &  $-0.09$     \\
        \ym &   & $\phantom{-}1$     &$-0.20$       &  $-0.05$     \\
        \xp &   &       &$ \phantom{-}1$          &  $\phantom{-}0.14$     \\
        \yp &   &       &            &  $\phantom{-}1$         \\
        \hline
        \end{tabularx}


\end{table}

\section{Results and interpretation}
\label{sec:Results}
The \CP observables are measured to be
\begin{align*}
    \xm & = (\phantom{-}9.0 \pm 1.7 \pm 0.7 \pm 0.4) \times 10^{-2}, \\
    \ym & = (\phantom{-}2.1 \pm 2.2 \pm 0.5 \pm 1.1) \times 10^{-2}, \\
    \xp & = (         - 7.7 \pm 1.9 \pm 0.7 \pm 0.4) \times 10^{-2}, \\
    \yp & = (         - 1.0 \pm 1.9 \pm 0.4 \pm 0.9) \times 10^{-2},
\end{align*}
where the first uncertainty is statistical, the second is the total experimental systematic uncertainty and the third is that arising from the precision of the CLEO measurements.

The signature for \CP violation is that $(\xp, \yp)\neq (\xm, \ym)$. The distance between $(\xp, \yp)$ and $(\xm, \ym)$ is calculated, taking all uncertainties and correlations into account, and found to be $|(\xp, \yp)-(\xm, \ym)| = (17.0\pm2.7)\times 10^{-2}$, which is different from zero by 6.4 standard deviations. This constitutes the first observation of \CP violation in \BtoDK decays for the $\D\to\KS h^+h^-$ final states.

\begin{figure}[t]
    \centering
    \includegraphics[width=.69\textwidth]{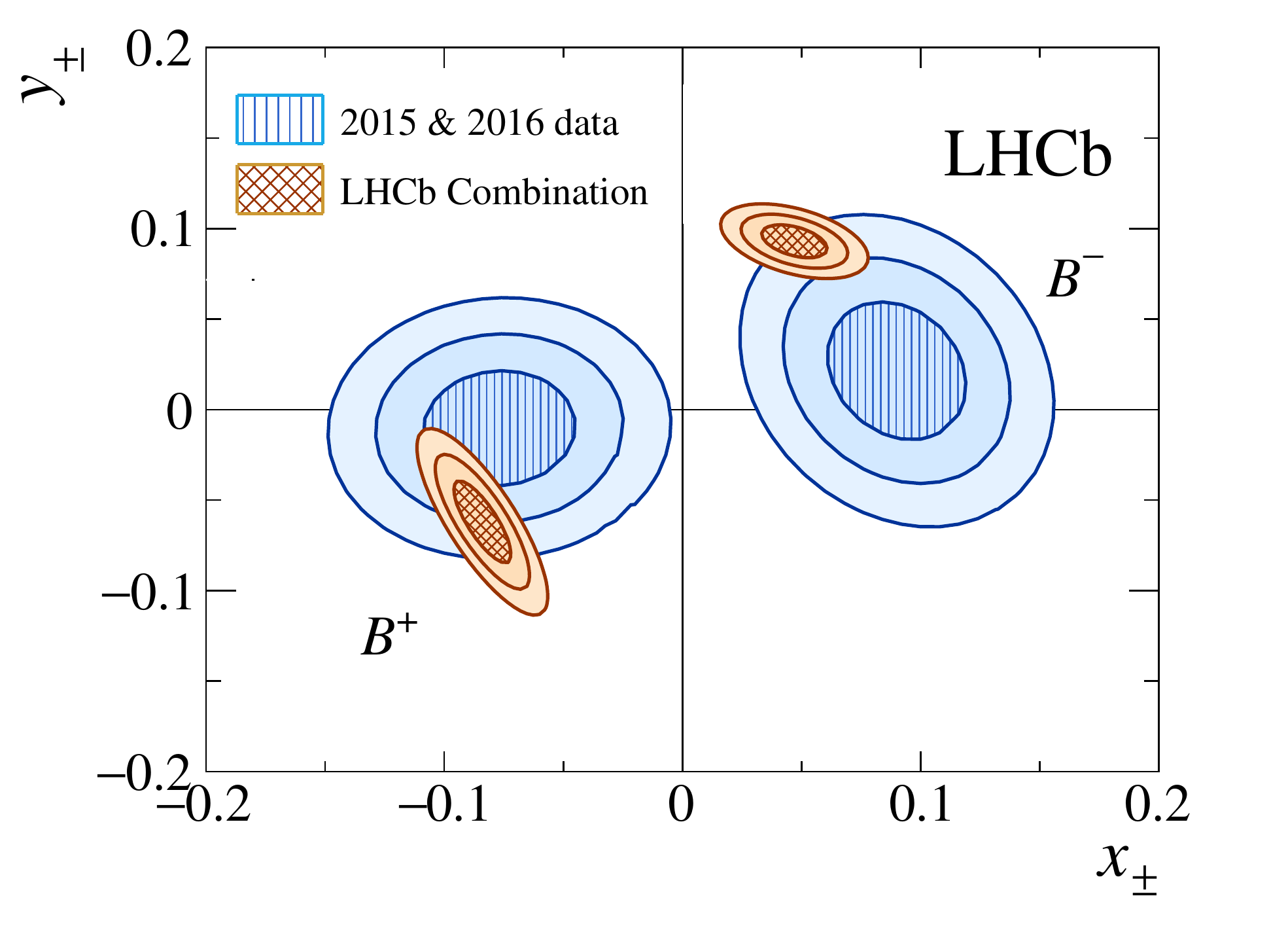}
    \caption{Two-dimensional 68.3\,\%, 95.5\,\% and 99.7\,\% confidence regions for $(\xpm,\, \ypm)$ obtained in this measurement, as well as for the LHCb combination in Ref.~\cite{LHCb-CONF-2017-004}, taking statistical and systematic uncertainties, as well as their correlations, into account.}
    \label{fig:xy_comparison}
\end{figure}

These results are compared to the expected central values of $x_\pm$ and $y_\pm$ that can be computed from $r_B, \delta_B$, and $\gamma$ as determined in the \lhcb combination in Ref.~\cite{LHCb-CONF-2017-004}, and the results are shown in Fig.~\ref{fig:xy_comparison} (the later \lhcb combination in Ref.~\cite{LHCb-CONF-2018-002} includes the results of this measurement and is therefore unsuitable for comparison).  The two sets of $(x_+,y_+)$ are in agreement within 1.6 standard deviations when the uncertainties and correlations of both the LHCb combination and this measurement are taken into account. There is a 2.7 standard deviation tension between the measured values of $(x_-,\,y_-)$ and the values calculated from the LHCb combination. This tension will be investigated further when this measurement and the LHCb combination are updated using data taken in 2017 and 2018.

The results for $x_\pm$ and $y_\pm$ are interpreted in terms of the underlying physics parameters $\gamma$, $r_B$ and $\delta_B$. The interpretation is done via a maximum likelihood fit using a frequentist treatment as described in Ref.~\cite{LHCb-PAPER-2016-032}. The solution for the physics parameters has a two-fold ambiguity as the equations are invariant under the simultaneous substitutions $\gamma\to\gamma+180^\circ$ and $\delta_B\to\delta_B+180^\circ$. The solution that satisfies $0<\gamma<180^\circ$ is chosen. The central values and 68\% (95\%) confidence intervals, calculated with the \textsc{PLUGIN} \cite{PLUGIN} method, are
\begin{align*}
    \gamma          &= 87^\circ \,  ^{+11^\circ}_{-12^\circ}\left(^{+22^\circ}_{-23^\circ}\right), \\
    r_B        &= 0.086   \,     ^{+0.013}   _{-0.014}   \left(^{+0.025}   _{-0.027}\right), \\
    \delta_B   &= 101^\circ \, ^{+11^\circ}_{-11^\circ}\left(^{+22^\circ}_{-23^\circ}\right).
\end{align*} 
The values for $\gamma$ and $r_B$ are consistent with those presented in Ref.~\cite{LHCb-CONF-2017-004}. This is the most precise measurement of $\gamma$ from a single analysis. The value of $\delta_B$ shows some disagreement with Ref.~\cite{LHCb-CONF-2017-004}, where the angle is determined to be $\left(139.9\,^{+4.8}_{-5.2}\right)^\circ$. 

The values of $x_\pm,y_\pm$ measured in this analysis can be combined with those from the corresponding analysis of Run 1 data~\cite{LHCb-PAPER-2014-041}. This procedure is done via  a maximum likelihood fit, as implemented in 
the \texttt{gammacombo} package \cite{LHCb-PAPER-2016-032}.
The previous measurements are identified by the index I, and the results within this paper are identified by the index II.
When combining the two results, the fit determines the $(\hat x_\pm, \hat y_\pm)$ parameters that maximize the multivariate Gaussian likelihood function
\begin{align}
    \mathcal{L}(z|\hat z) = ((2\pi)^8|\Sigma|)^{-1/2}\exp\left[-\frac{1}{2}(z-\hat z)^T\Sigma^{-1} (z-\hat z)\right],
\end{align}
where $z=(\xpm^{\text{I}}, \ypm^{\text{I}}, \xpm^{\text{II}}, \ypm^{\text{II}})^T$ 
and $\hat z = (\hat x_\pm, \hat y_\pm, \hat x_\pm, \hat y_\pm)^T $ are $8\times 1$ 
vectors and $\Sigma$ is the $8\times8$ covariance matrix
\begin{align}
    \Sigma = \left(
    \begin{array}{cc}
        \Sigma^\text{I} & \Sigma^\text{I--II}\\
        \Sigma^\text{II--I} & \Sigma^\text{II}
    \end{array}
    \right).
\end{align}
The covariance matrix is expressed in terms of the covariance matrices obtained for the individual measurements, 
$\Sigma^\text{I}$ and $\Sigma^\text{II}$, and the cross-covariance matrix $\Sigma^\text{I--II}$ describing 
correlations between the measurements. The covariance matrix for this measurement, $\Sigma^\text{II}$, is calculated using the total statistical and systematic uncertainties, and the correlation matrices in Tables~\ref{tab:stat_correlations}~and~\ref{tab:syst_correlations}. The covariance matrix for the Run~1 measurement, $\Sigma^\text{I}$, is taken from Ref.~\cite{LHCb-PAPER-2014-041}, where it was calculated taking strong-phase-related correlations into account, but treating the experimental systematic uncertainties as uncorrelated. The impact of using the correlation matrix in Table~\ref{tab:syst_correlations} for these instead is found to be negligible.

The dominant uncertainty in both measurements is the statistical uncertainty. As the measurements use independent data sets, the
statistical uncertainties are uncorrelated.
The cross-correlations of the systematic errors between measurements due to the strong phase inputs are obtained from the results 
of a series of fits to the two data sets in which the strong phases are varied identically.
This mirrors the procedure used to evaluate the uncertainties within a single data set.
The obtained cross-correlations between the fit results are given in Table~\ref{tab:cleo_cross_correlations}. 
The elements on the diagonal do not have unit value because the obtained correlations depend on the specific data sets for the two measurements.

\begin{table}[t]
    \renewcommand\arraystretch{1.2}
    \centering

    \caption{
    Correlation matrix between Run~1 results (I) and the results presented in this paper~(II), when fitting data while varying the inputs from the CLEO collaboration in a correlated way.
    \label{tab:cleo_cross_correlations}
    }

        \begin{tabularx}{0.6\textwidth}{l|YYYY}
        \hline
        \multicolumn{5}{c}{CLEO cross-run correlation matrix} \\
        \hline
        & $\phantom{-}\xm^{\text{II}}$ & $\phantom{-}\ym^{\text{II}}$ &$ \phantom{-}\xp^{\text{II}}$ &$ \phantom{-}\yp^{\text{II}}$ \\
        \hline
        $\xm^{\text{I}}$  & $\phantom{-}0.02 $ & $\phantom{-}0.35  $ & $-0.32 $&  $-0.21$  \\
        $\ym^{\text{I}}$  & $\phantom{-}0.45 $ & $-0.23 $ & $ \phantom{-}0.03 $&  $-0.28$  \\
        $\xp^{\text{I}}$  & $-0.19$ & $\phantom{-} 0.01 $ & $ \phantom{-}0.55 $&  $-0.22$ \\
        $\yp^{\text{I}}$  & $-0.30$ & $ -0.28$ & $  \phantom{-}0.13$&  $ \phantom{-}0.48$    \\
        \hline
        \end{tabularx}

\end{table}
\begin{table}[t]
    \centering
    \renewcommand\arraystretch{1.2}
    \caption{
    Total correlation matrix for the systematic uncertainties of the Run~1 results (I) and the results presented in this paper (II), including experimental and strong phase related systematic uncertainties.
    \label{tab:total_cross_correlations}
    }

        \begin{tabularx}{0.6\textwidth}{l|YYYY}
        \hline
        \multicolumn{5}{c}{Total systematic cross-run correlation matrix} \\
        \hline
        & $\phantom{-}\xm^{\text{II}}$ & $\phantom{-}\ym^{\text{II}}$ &$ \phantom{-}\xp^{\text{II}}$ &$ \phantom{-}\yp^{\text{II}}$ \\
        \hline
        $\xm^{\text{I}}$  &$ \phantom{-}0.76$ & $ \phantom{-}0.04$ & $ \phantom{-}0.55$ & $ \phantom{-}0.02$ \\
        $\ym^{\text{I}}$  &$ \phantom{-}0.14$ & $-0.06$ & $-0.13$ & $-0.25$ \\
        $\xp^{\text{I}}$  &$ \phantom{-}0.58$ & $-0.19$ & $ \phantom{-}0.91$ & $ \phantom{-}0.05$ \\
        $\yp^{\text{I}}$  &$-0.05$ & $-0.24$ & $ \phantom{-}0.17$ & $ \phantom{-}0.55$ \\
        \hline
        \end{tabularx}

\end{table}

\begin{figure}[t]
    \centering
    \includegraphics[width=0.46\textwidth]{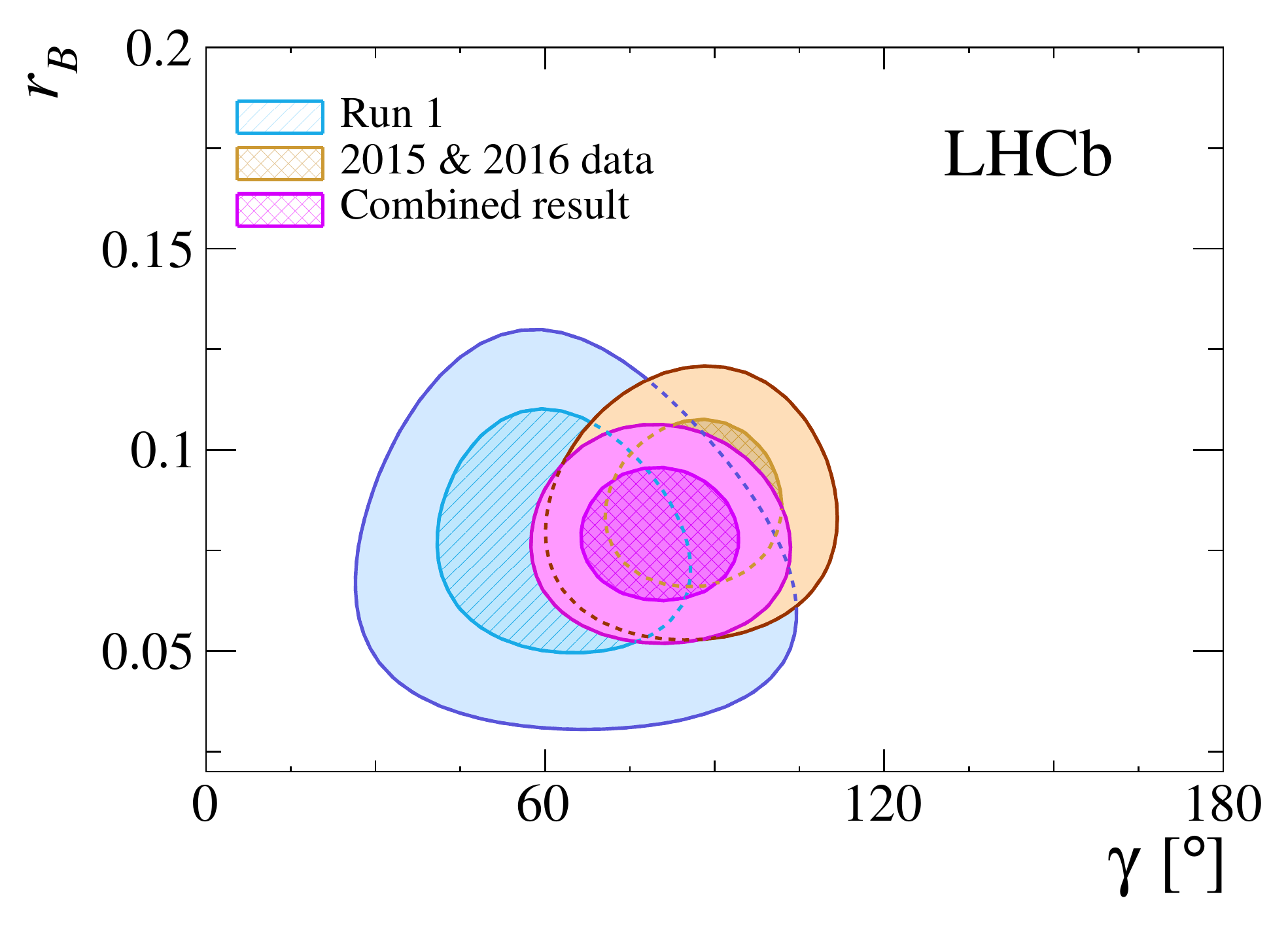}
    \includegraphics[width=0.46\textwidth]{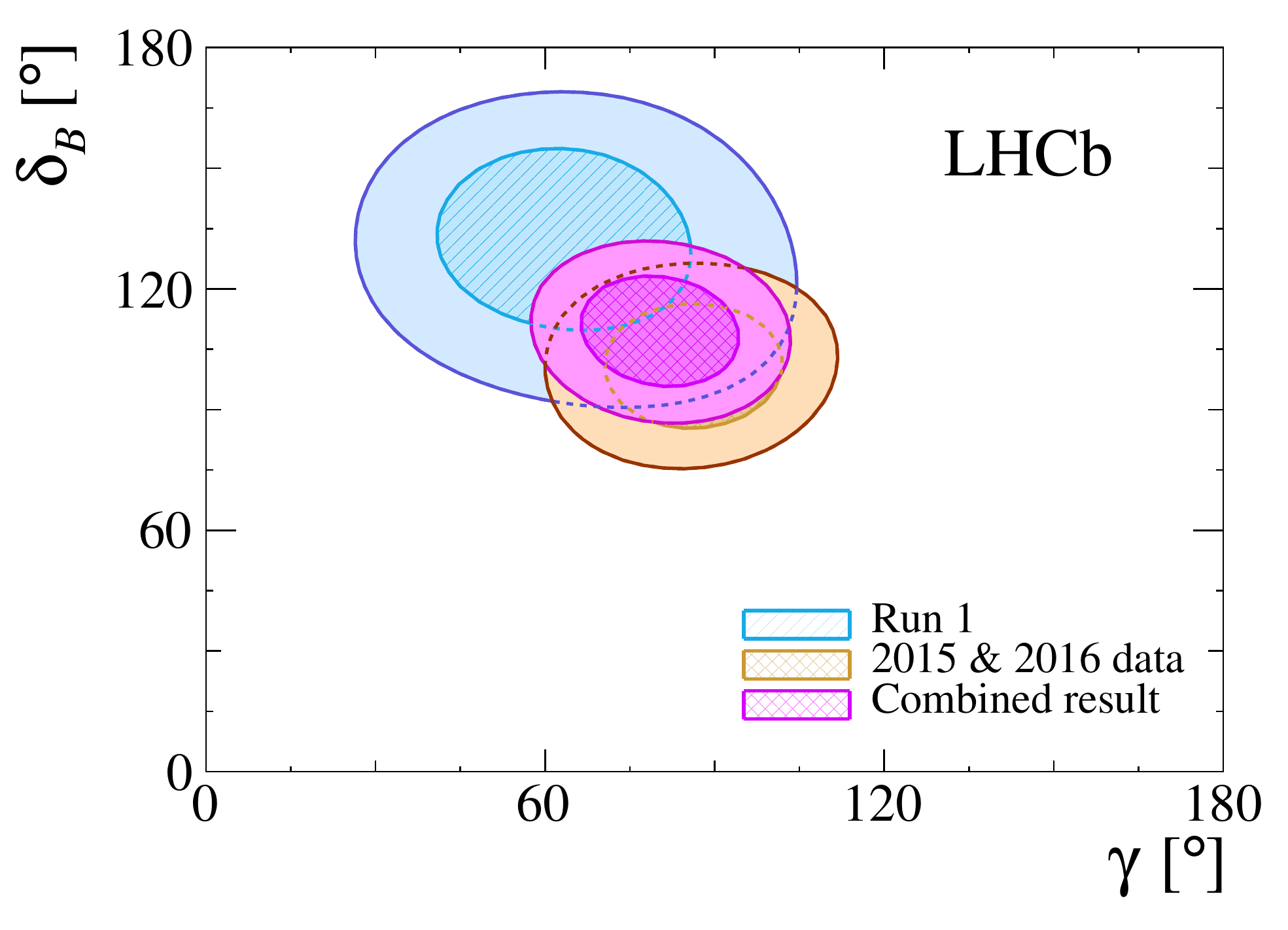}
    \caption{Two-dimensional 68.3\,\% and 95.5\,\% confidence regions for $( \gamma,  r_B,  \delta_B)$ for the \xy parameters obtained in the fit to 2015 and 2016 data, the fit to Run~1 data, and their combinations. }
    \label{fig:fig_2d_run12}
\end{figure}

The combination is performed assuming full correlation between the non-strong-phase related experimental systematic uncertainties in Run~1 and this measurement. The correlation matrix for the experimental uncertainties of this analysis is used as the cross-run correlation of the experimental systematic uncertainties. The complete correlation matrix for the experimental and strong-phase-related systematic uncertainties is given in Table~\ref{tab:total_cross_correlations}. The impact on the combination due to different assumptions on the cross-correlations of the systematic uncertainties is found to be negligible. This is unsurprising as both measurements remain limited in precision by their statistical uncertainties. 
The central values, along with the combined statistical 
and systematic uncertainties for this combination are
\begin{align*}
    <\xm> & =    (\phantom{-}7.0 \pm 1.7 ) \times 10^{-2}, \\
    <\ym> & =    (\phantom{-}4.1 \pm 2.0 ) \times 10^{-2}, \\
    <\xp> & =    (         - 7.8 \pm 1.7 ) \times 10^{-2}, \\
    <\yp> & =    (         - 1.4 \pm 1.7 ) \times 10^{-2}.
\end{align*}
The interpretation in terms of the underlying physics parameters is performed on the combined values of $x_\pm$ and $y_\pm$ and the central values and their 68$\%$ (95$\%$) confidence intervals are 
\begin{align*}
    \gamma          &= 80^\circ \,  ^{+10^\circ}_{-9^\circ}\left(^{+19^\circ}_{-18^\circ}\right), \\
    r_B        &= 0.080  \,      ^{+0.011}   _{-0.011}  \left (^{+0.022}   _{-0.023}\right), \\
    \delta_B   &= 110^\circ \, ^{+10^\circ}_{-10^\circ}\left(^{+19^\circ}_{-20^\circ}\right). 
\end{align*}
The results of the interpretation for both the combined and individual data sets are shown in Fig.~\ref{fig:fig_2d_run12}, where the projections of the three-dimensional surfaces bounding the one and two standard deviation volumes on the ($\gamma$, $r_B$) and ($\gamma$, $\delta_B$) planes are shown. 
The uncertainty on $\gamma$ is inversely proportional to $r_B$. Therefore the lower central value of $r_B$ in the combined results lead to a larger than naively expected uncertainty on $\gamma$ when both data sets are used. The contribution of each source of uncertainty are estimated by performing the combination while taking only subsets of the uncertainties into account. It is found that the statistical uncertainty on $\gamma$ is $8.5^\circ$, the uncertainty due to strong-phase inputs is $4^\circ$, and the uncertainty due to experimental systematic effects is $2^\circ$.

\section{Conclusions}
\label{sec:conclustions}

Approximately 4100 (560) $B^\pm \to D K^\pm$ decays with the $D$ meson decaying to $\KS \pi^+\pi^-$ ($\KS K^+ K^-$) are selected from data corresponding to an integrated luminosity of 2.0~${\rm fb^{-1}}$ collected with the LHCb detector in 2015 and 2016. These samples are analysed to determine the \CP-violating parameters $x_\pm \equiv r_B \cos (\delta_B \pm \gamma)$ and $y_\pm \equiv r_B \sin (\delta_B \pm \gamma)$, where $r_B$ is the ratio of the absolute values of the $B^+ \to D^0 K^-$ and $B^+ \to \Dzb K^-$ amplitudes,  $\delta_B$ is their strong-phase differences, and $\gamma$ is an angle of the Unitarity Triangle. 
The analysis is performed in bins of the $D$-decay Dalitz plot and existing measurements performed by the CLEO collaboration~\cite{CLEOCISI} are used to provide input on the $D$-decay strong-phase parameters $(c_i, s_i)$. Such an approach allows the analysis to be free from model-dependent assumptions on the strong-phase variation across the Dalitz plot. This paper also gives the combination with the results obtained with an earlier data set, thereby allowing further improvements in the precision on $\gamma$. Considering only the data collected in 2015 and 2016 and choosing the solution that satisfies $0<\gamma<180^\circ$ yields \linebreak $r_B$ = $0.086^{+0.013}_{-0.014}$, $\delta_B = (101 \pm 11)^\circ$, and $\gamma = (87^{+11}_{-12})^\circ$. The values of $r_B$ and $\gamma$ are consistent with world averages, while there is some tension in the determined value of $\delta_B$. This could be resolved by future analyses of the $B \to D K$ mode in a variety of $D$ decays, including those analysed here, utilising the data set that is being collected with LHCb in 2017 and 2018. The measurement reported in this paper represents the most precise determination of $\gamma$ from a single analysis.




\section*{Acknowledgements}
%
%
\noindent We express our gratitude to our colleagues in the CERN
accelerator departments for the excellent performance of the LHC. We
thank the technical and administrative staff at the LHCb
institutes. We acknowledge support from CERN and from the national
agencies: CAPES, CNPq, FAPERJ and FINEP (Brazil); MOST and NSFC
(China); CNRS/IN2P3 (France); BMBF, DFG and MPG (Germany); INFN
(Italy); NWO (Netherlands); MNiSW and NCN (Poland); MEN/IFA
(Romania); MinES and FASO (Russia); MinECo (Spain); SNSF and SER
(Switzerland); NASU (Ukraine); STFC (United Kingdom); NSF (USA).  We
acknowledge the computing resources that are provided by CERN, IN2P3
(France), KIT and DESY (Germany), INFN (Italy), SURF (Netherlands),
PIC (Spain), GridPP (United Kingdom), RRCKI and Yandex
LLC (Russia), CSCS (Switzerland), IFIN-HH (Romania), CBPF (Brazil),
PL-GRID (Poland) and OSC (USA). We are indebted to the communities
behind the multiple open-source software packages on which we depend.
Individual groups or members have received support from AvH Foundation
(Germany), EPLANET, Marie Sk\l{}odowska-Curie Actions and ERC
(European Union), ANR, Labex P2IO and OCEVU, and R\'{e}gion
Auvergne-Rh\^{o}ne-Alpes (France), Key Research Program of Frontier
Sciences of CAS, CAS PIFI, and the Thousand Talents Program (China),
RFBR, RSF and Yandex LLC (Russia), GVA, XuntaGal and GENCAT (Spain),
Herchel Smith Fund, the Royal Society, the English-Speaking Union and
the Leverhulme Trust (United Kingdom).



\addcontentsline{toc}{section}{References}
\setboolean{inbibliography}{true}
\bibliographystyle{LHCb}
\bibliography{main,LHCb-PAPER,LHCb-CONF,LHCb-DP,LHCb-TDR}

\newpage


\newpage
\centerline{\large\bf LHCb collaboration}
\begin{flushleft}
\small
R.~Aaij$^{27}$,
B.~Adeva$^{41}$,
M.~Adinolfi$^{48}$,
C.A.~Aidala$^{73}$,
Z.~Ajaltouni$^{5}$,
S.~Akar$^{59}$,
P.~Albicocco$^{18}$,
J.~Albrecht$^{10}$,
F.~Alessio$^{42}$,
M.~Alexander$^{53}$
% ...

{\footnotesize \it
% ...
$ ^{5}$Clermont Universit{\'e}, Universit{\'e} Blaise Pascal, CNRS/IN2P3, LPC, Clermont-Ferrand, France\\
% ...
$ ^{10}$Fakult{\"a}t Physik, Technische Universit{\"a}t Dortmund, Dortmund, Germany\\
% ...
$ ^{18}$INFN Laboratori Nazionali di Frascati, Frascati, Italy\\
% ...
$ ^{27}$Nikhef National Institute for Subatomic Physics, Amsterdam, Netherlands\\
% ...
$ ^{41}$Instituto Galego de F{\'\i}sica de Altas Enerx{\'\i}as (IGFAE), Universidade de Santiago de Compostela, Santiago de Compostela, Spain\\
$ ^{42}$European Organization for Nuclear Research (CERN), Geneva, Switzerland\\
% ...
$ ^{48}$H.H. Wills Physics Laboratory, University of Bristol, Bristol, United Kingdom\\
% ...
$ ^{53}$School of Physics and Astronomy, University of Glasgow, Glasgow, United Kingdom\\
% ...
$ ^{59}$University of Cincinnati, Cincinnati, OH, United States\\
% ...
$ ^{61}$Syracuse University, Syracuse, NY, United States\\
% ...
$ ^{73}$University of Michigan, Ann Arbor, United States, associated to $^{61}$\\
% ...
}
\end{flushleft}

\end{document}